\documentclass[reprint,amsmath,amssymb,aps,pre,showpacs,floatfix,footinbib]{revtex4-1}
\usepackage{lineno}
\usepackage{xcolor,soul}
\soulregister\cite7
\soulregister\eqref7
\usepackage{verbatim,amsmath,mathtools,nicefrac,amssymb,graphicx,color}
\usepackage{epsfig,epstopdf,url,braket}

\newcommand{\kv}{\mathbf{k} }

\begin{document}
\title{The sound speed in Yukawa one component plasmas across coupling regimes}

\author{Luciano G. Silvestri}
\altaffiliation{Present address: Computational Mathematics Science and Engineering, Michigan State University, East Lansing, MI, 48824, USA}
\affiliation{Department of Physics, Boston College, Chestnut Hill, MA, USA}
\author{Gabor J. Kalman}
\affiliation{Department of Physics, Boston College, Chestnut Hill, MA, USA}
\author{Zolt\'an Donk\'o}
\affiliation{Institute of Solid State Physics and Optics, Wigner Research Centre for Physics, Budapest, HU}
\author{Peter Hartmann}
\affiliation{Institute of Solid State Physics and Optics, Wigner Research Centre for Physics, Budapest, HU}
\author{Marlene Rosenberg}
\affiliation{Department of Electrical and Computer Engineering, University of California San Diego, San Diego, CA, USA}
\author{Kenneth I. Golden}
\affiliation{College of Engineering and Mathematical Sciences, University of Vermont, Burlington, VT, USA}
\author{Stamatios Kyrkos}
\affiliation{Department of Physics, Le Moyne College, Syracuse, NY, USA}

\begin{abstract}
A many-body system of charged particles interacting via a pairwise Yukawa potential, the so-called Yukawa One Component Plasma (YOCP) is a good approximation for a variety of physical systems. Such systems are completely characterized by two parameters; the screening parameter, $\kappa$, and the nominal coupling strength, $\Gamma$. It is well known that the collective spectrum of the YOCP is governed by a longitudinal acoustic mode, both in the weakly and strongly coupled regimes. In the long-wavelength limit the linear term in the dispersion (\textit{i.e.} $\omega = s k$) defines the sound speed $s$. We study the evolution of this latter quantity from the weak through the strong coupling regimes by analyzing the Dynamic Structure Function $S(\kv,\omega)$ in the low frequency domain.  Depending on the values of $\Gamma$ and $\kappa$ and $w = s/v_{\textrm{th}}$, (\textit{i.e.} the ratio between the phase velocity of the wave and thermal speed of the particles) we identify five domains in the $(\kappa,\Gamma)$ parameter space in which the physical behavior of the YOCP exhibits different features. The competing physical processes are the collective Coulomb like vs. binary collision dominated behavior and the individual particle motion vs. quasi-localization. Our principal tool of investigation is  Molecular Dynamics (MD) computer simulation from which we obtain $S(\kv,\omega)$. Recent improvements in the simulation technique have allowed us to obtain a large body of high quality data in the range $\Gamma = \{0.1 - 10,000\}$ and $\kappa = \{0.5 - 5\}$. The theoretical results based on various models are compared in order to see which one provides the most cogent physical description and the best agreement with MD data in the different domains.
\end{abstract}
\date{\today}
\maketitle
A many body system of charged particles interacting via a pairwise Yukawa potential, the so-called Yukawa One Component plasma (YOCP), is considered to be a good approximation for a large variety of strongly coupled systems. Examples include dusty plasmas \cite{Morfill2009}, warm dense matter \cite{Murillo2010}, and ultracold plasmas \cite{Killian2007}. Charges in these system interact via the Yukawa pair potential of the form
\begin{equation}
    \phi(r) = \frac{q^2}{r}\exp[{-r/\lambda}]
    \label{yr}
\end{equation}
or in wave number $\mathbf k$ space:
\begin{equation}
    \tilde \phi(k) = \frac{q^2}{k^2 + \lambda^{-2}},
\end{equation}
where $q$ is the charge of the particles, $\lambda$ is the screening length, and $k = |\kv|$. The YOCP is characterized by two dimensionless parameters: the coupling strength $\Gamma$ and the screening parameter $\kappa$ defined as
\begin{equation}
    \Gamma = \frac{\beta q^2}{a}, \quad \kappa = a/\lambda 
    \label{eq:YOCPparam}
\end{equation} 
where $a = (3/4\pi n)^{1/3} $ is the Wigner-Seitz radius, $n$ the density, and $\beta = 1/k_BT$ the inverse temperature. The dynamical structure function (DSF) $S(\kv,\omega)$ of a strongly coupled YOCP has been extensively studied in the literature. Salin derived an expression for $S(\kv,\omega)$ from linearized hydrodynamics equations and showed that it is given by the sum of three Lorentzians centered at $\omega = 0$, $\omega = \pm c_sk$, representing a diffusive mode and two propagating acoustic modes with sound speed $c_s$, respectively \cite{Salin2007}.  Mithen \textit{et al.} have concluded that such description well reproduces the DSF of YOCP for wave numbers $ ka < 0.43 \kappa $, while at higher wave numbers a more sophisticated approach, based perhaps on memory functions, is needed \cite{Mithen2011a,Mithen2011b}. A more precise description of the collective mode spectrum of the YOCP is, by now, available \cite{Rosenberg1997,Kaw1998,KalmanDeWitt2000,Ohta2000,Murillo2000}. Both the weakly ($\Gamma \ll 1$) and strongly coupled regimes ($\Gamma \gg 1$) are governed by a longitudinal acoustic mode that becomes the longitudinal plasmon, by means of the Anderson-like mechanism, in the $\kappa = 0$ limit \cite{Anderson1963}. In the small $k$ region the acoustic mode has a linear dispersion, \textit{i.e.} $\omega = sk$ where $s$ is the longitudinal sound speed.  As to  the transverse sound which is maintained by correlations, no  such a  mode is present in the weakly coupled regime. A doubly degenerate
acoustic mode, with sound speed $s_{\text{T}} < s$, 	emerges only for $\Gamma \ge 50$ \cite{KalmanDeWitt2000,Mithen2014}. Furthermore, the theoretical prediction that the transverse mode extends all the way to $\kv = 0$ is spurious, since the liquid is unable to support shear waves in the uniform limit \cite{Donko2008}. Hence the focus of this paper is the longitudinal sound, since its occurrence across the entire coupling regime makes it amenable to the study of how the different phases of the YOCP affect its phase velocity. 

In order to calculate the sound speed a variety of different theoretical methods have been used in the literature:  generalized hydrodynamics \cite{Kaw1998}, static and dynamic local field correction of the response function \cite{Murillo2000}, Sum Rule based Feynmann approach \cite{Murillo2010,Diaw2015}, and Quasi-Localized Charge Approximation \cite{Rosenberg1997,KalmanDeWitt2000,Rosenberg2016}. Recently, Khrapak and Thomas \cite{Khrapak2015b,*Khrapak2016} have shown how a simple fluid description complemented with a sufficiently general phenomenological equation of state valid across coupling regimes is sufficient to provide a good estimate for the sound speed. 

All of the above research, however, has focused on the strongly coupled regime. In this work we attempt to describe the behavior of the sound speed and damping of the oscillations across a wide range of parameter regimes: from a weakly coupled YOCP to its  crystalline phase. We will also indicate what the appropriate theoretical models for these different domains seem to be. Our principal criterion in assessing their merits is internal consistency and agreement with Molecular Dynamics (MD) simulations. We also provide analytic calculations through different theoretical models, with our emphasis being on the QLCA, which has proven to be a highly reliable predictor in the strongly and moderately coupled regimes \cite{Rosenberg1997,Rosenberg2016}. 
Comparison with recent MD data will help determine the coupling value down to which the QLCA and its underlying model are applicable.

\begin{figure*}[ht]
\centering
\includegraphics[width = \linewidth]{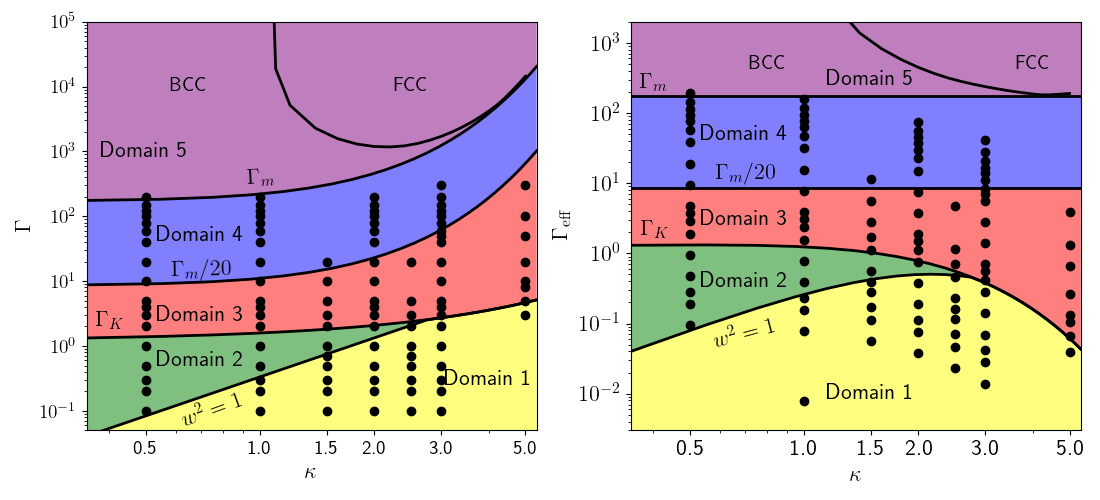}
\caption{(Left) $\Gamma-\kappa$ parameter space. (Right) $\Gamma_{\textrm{eff}} -\kappa$ parameter space. The boundaries are defined in the text. The black dots indicate all the simulations performed.}
\label{domainsplot}
\end{figure*}

\begin{figure*}[ht]
    \centering
    \includegraphics[width = \linewidth]{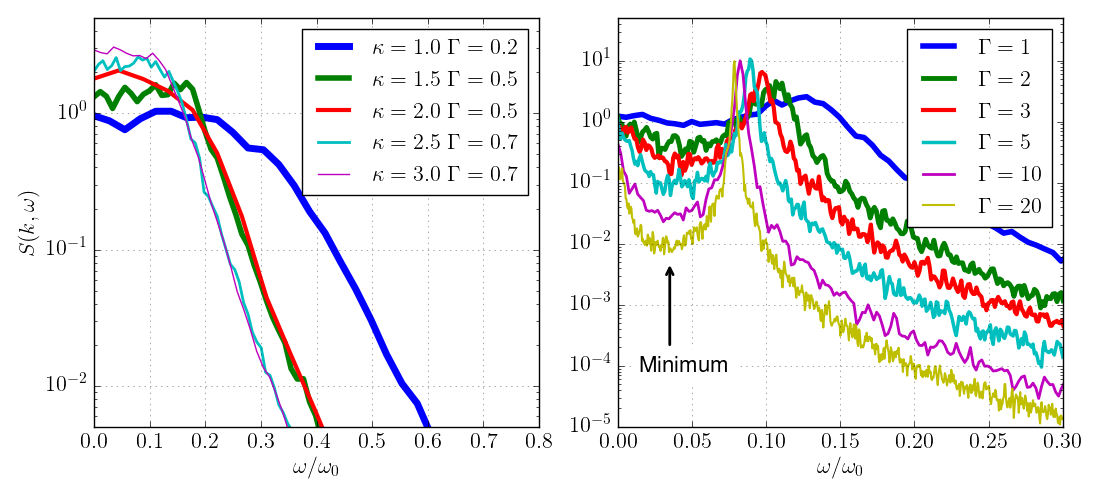}
    \caption{(Color online) Plots of $S(\kv,\omega)$ obtained from MD simulation at the lowest $ka$ value. (Left) $\Gamma-\kappa$ values pertaining to Domain 1. (Right) Evolution of the minimum in $S(\kv,\omega)$ as the system enters Domain 3.}
    \label{domain2plot}
\end{figure*}

\section{Computer Simulations}
\label{sec:simulations}
Our MD simulations have been done for a system of $N = 27\, 648$ point--like charged particles with charge $q$ and mass $m$ interacting via the Yukawa potential \eqref{yr} in a cubic box of side $L$, to which periodic boundary conditions are applied. The finite length of the cube restricts the lowest possible value of the wave vector to
\begin{equation}
    ka_{\textrm {min} } = \frac{2\pi}L a = \frac{2\pi}{L} \left( \frac{3 L^3}{4\pi N} \right)^{1/3} = 0.1289
\end{equation}
Recent improvements in the simulation technique have allowed us to obtain a large body of high quality data in the range $ \Gamma = \{0.1 - 10\,000 \}$ and $\kappa = \{ 0.5 - 5 \}$. Upon initialization the particles are placed randomly inside the simulation box and their velocities are sampled from a Maxwell-Boltzmann distribution having a temperature that corresponds to the pre-defined $\Gamma$. The equations of motion of the particles are integrated with the Velocity-Verlet scheme with a time step $\lesssim 1/(20 \omega_0)$ where $\omega_0$ is the plasma frequency defined in eq.~\eqref{speedc0}. For simulations with $\Gamma < 1$ a time-step $\lesssim a/(50 v_{\textrm{th}} )$ was chosen, in order to correctly resolve ballistic motion. Following the initialization of the simulations, the thermalization of the system is achieved by rescaling the velocities of the particles at each time step until equilibration. During this thermalization phase no measurements are taken on the system. Once the system has reached equilibrium, the positions and velocities of the particles are recorded; from these data we obtain the microscopic density \cite{Hansen1975} 
\begin{equation}
    n_{\mathbf k}(t) = \sum_j e^{i \mathbf k\cdot \mathbf x_j(t)}
\end{equation}
and the DSF $S(\kv,\omega)$
\begin{equation}
    S(\kv,\omega) = \frac{1}{2\pi N} \frac{1}{\Delta \tau}\left | n_{\mathbf k}(\omega) \right |^2
\end{equation}
where $n_{\mathbf k}(\omega)$ is the Fourier transform of $n_{\mathbf k}(t)$ and $\Delta \tau $ is the length of the data recording period typically $\omega_0 \Delta \tau  \sim 500 - 1000$. The peak of $S(\kv,\omega)$ identifies the collective mode, from whose position we calculate the sound speed by fitting an appropriate function to $S(\kv,\omega)$ and subsequently fitting the peak positions at the lowest three $ka$ values to a function of the form $f(x) = ax + bx^3$. The coefficient of the linear term is then identified as the sound speed. In addition to the dynamical structure function, we calculate the static structure function $S(\kv)$ and the pair distribution function $g(r)$ as these serves as the basis of the theoretical calculations of the sound speed.

\section{Domains}
\label{sec:domains}
A major result of this work is the identification of five different domains in the $\Gamma-\kappa$ parameter space. In each of these Domains the physical behavior of the YOCP exhibits different features. Fig.~\ref{domainsplot} shows two plots of the parameter space divided into the five domains. The left hand side plot represents a parameter space spanned by $\kappa$ and $\Gamma$, while the right hand side one the parameter space spanned by $\kappa$ and $\Gamma_{\textrm{eff}}$, an effective coupling strength defined by \cite{Vaulina2002,Ott2014,Khrapak2015b,*Khrapak2016}
\begin{equation}
    \Gamma_{\textrm{eff}}(\kappa,\Gamma) = \Gamma \left ( 1 + \alpha \kappa + \alpha^2\kappa^2/2 \right ) e^{-\alpha \kappa}, \; \alpha = (4\pi/3)^{1/3}.
    \label{geff}
\end{equation}
We note that Ott \textit{et al.} \cite{Ott2014} proposed a different definition of  $\Gamma_{\textrm{eff}}$. Their results are, however, valid only  in the range $1 \leq \Gamma \leq 150$  and $\kappa \leq 2$, which is more restricted than the one considered here. At weak coupling we introduce the additional parameter:
\begin{equation}
	w = \frac{c_0}{v_{\textrm{th}} } = \sqrt{\frac{3\Gamma}{\kappa^2} }.
	\label{wratio}
\end{equation}
It represents the ratio of the nominal phase velocity $c_0$ obtained from the Random Phase Approximation (RPA)
\begin{equation}
    c_0 = \omega_0a/\kappa, \quad \omega_0^2 = \frac{4\pi q^2n}{m}.
    \label{speedc0}
\end{equation}
and the thermal speed $v_{\textrm{th}} = (\beta m)^{-1/2}$ \footnote{$c_0$ is the conventional dust acoustic speed in a dusty plasma}. This parameter divides the weak coupling regime into two Domains, which are bounded from above by the Kirkwood line, given as
\begin{equation}
    \Gamma_{\textrm{K} } (\kappa) = 1.21e^{\kappa/3.7}
    \label{GKeq}
\end{equation}
separating regions where the correlation function $h(r) = g(r) - 1$ decays monotonically as $r\rightarrow \infty$ from those with an oscillating $h(r)$ \cite{Murillo2010,Hopkins2005}. In Domain 1, where $\Gamma < \Gamma_{\textrm{K}}$ and $w < 1$, the large thermal motion of the particles inhibits the formation  of a collective mode, and no acoustic peak is discernible in  $S(\kv,\omega$). This behavior is reported in plots of $S(\kv,\omega)$ for different pairs of ($\kappa,\Gamma$) in the left panel of Fig.~\ref{domain2plot}. Moving to Domain 2, characterized by $\Gamma < \Gamma_{\textrm{K}}$ and $w > 1$, the system behaves as a weakly coupled Coulomb gas. In Domain 3, $\Gamma_{\textrm{K}} < \Gamma < \Gamma_m/20$, the system is a moderately coupled liquid, with the melting line $\Gamma_m (\kappa)$ defined as
\begin{equation}
    \Gamma_m (\kappa)= \frac{\Gamma^{(\textrm{OCP})}_{m} e^{\alpha \kappa}}{1 + \alpha \kappa + \alpha^2 \kappa^2/2}.
    \label{Gm}
\end{equation}
where $\Gamma^{(\textrm{OCP})}_{m} = 172 $ is the melting value of the Coulomb OCP \cite{Stringfellow1990}. In this Domain the correlational contribution to the compressibility of the system becomes negative, leading to the creation of a minimum (or valley) in $S(\kv,\omega)$ at low frequencies, see right panel of  Fig.~\ref{domain2plot}. The upper boundary of Domain 3 has been heuristically  identified  as $\Gamma_m/20$. Continuing to Domain 4, $\Gamma > \Gamma_m/20$, the system enters the strongly coupled liquid phase. Finally, Domain 5 is  defined by values of $\Gamma > \Gamma_m$, and it represents the crystalline phase. This latter can be further divided into two subdomains, in which the system crystallizes either in a BCC or in a FCC structure, depending on the value of the screening parameter~\cite{Hamaguchi1997}. 

We note that for the case of ``complex'' (dusty) plasmas, an alternative diagrammatic classification has been presented in Ref.~\cite{Khrapak2004}. In this work the authors use a criterion different from ours to distinguish various phases, namely the scattering cross section for binary collisions. This allows them to  divide the entire phase diagram in two main regions: a weak coupling ``ideal'' regime, where the dust grains interact primarily via binary collisions, and a non-ideal regime, where many-body interactions are dominant. The work presented here differs from that of Ref.~\cite{Khrapak2004} inasmuch as our criterion for distinguishing the various phases is the evolution of correlations, whose effect on the longitudinal sound speed is of prime importance.
\section{Theoretical Models}
\label{sec:theory}
\begin{figure*}[ht]
    \centering
    \includegraphics[width = \linewidth]{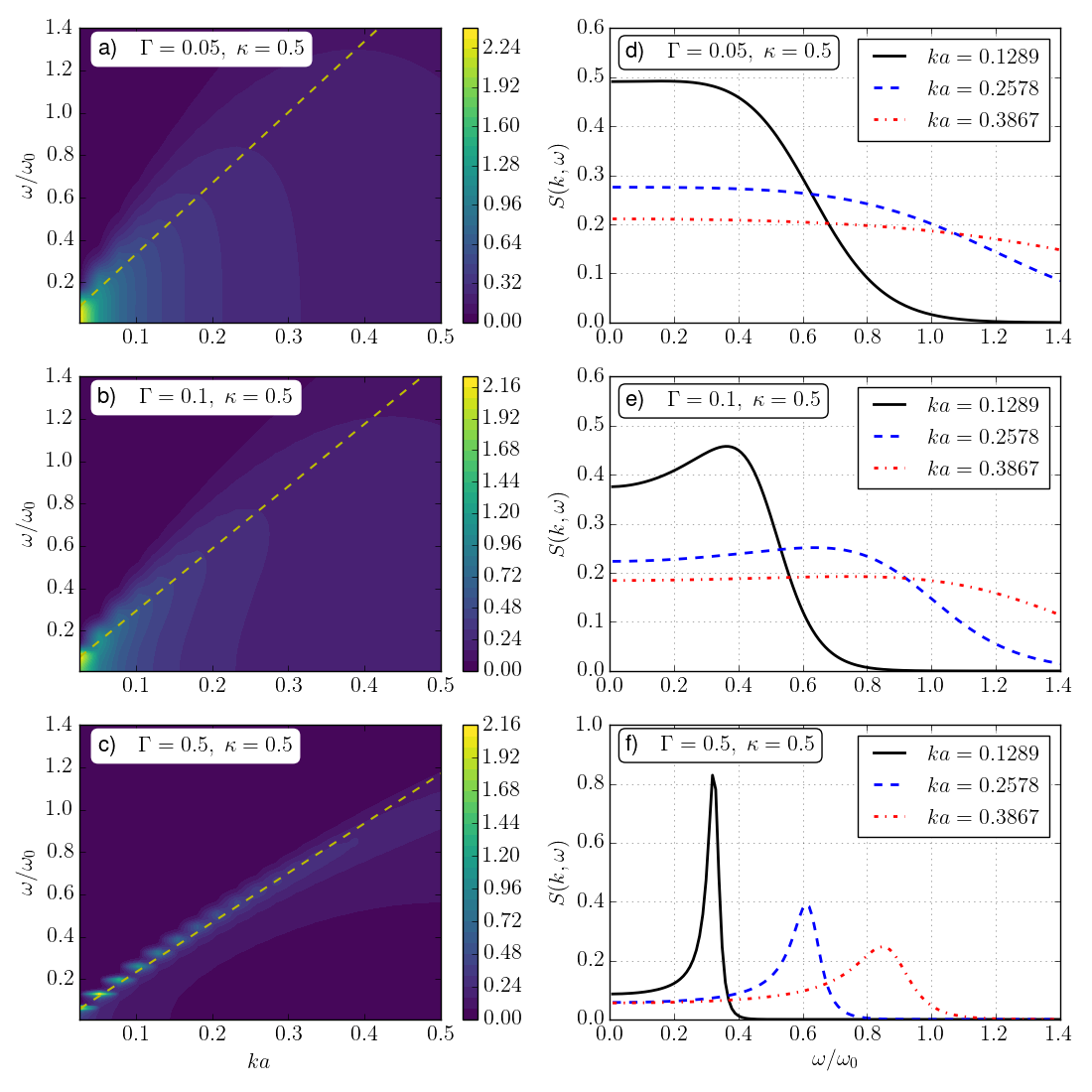}
    \caption{(Color online) Panels a) - c): color maps of $S(\kv,\omega)$, calculated using eqs.~\eqref{eq:chi0} and \eqref{eq:FDT}, showing the dispersion (or lack of) of the longitudinal acoustic mode, the dashed lines indicate the acoustic mode $\omega(k) = s_{\textrm{RPA}}k$ where $s_{\textrm{RPA}}$ is given by eq.~\eqref{eq:gksound}. Panels d) - f): line plots of $S(\kv,\omega)$ at three $ka$ values showing the presence (or lack of) of the acoustic peak. The $ka$ values are the same for all plots and indicate in the bottom one. The $\Gamma,\kappa$ values are indicated in the inserts. Panels a) and d) belong to Domain 1, b) and e) on the boundary between Domain 1 and Domain 2, and c) and f) plots belong to Domain 2.}
    \label{RPAclrplotsk05}
\end{figure*}
The dynamics of Yukawa systems is governed by the competition of two physical effects: the long range collective behavior as inherited from the progenitor of the YOCP, the simple Coulomb OCP (COCP), and short range binary interactions, more akin to those taking place in a neutral gas of atoms. Based on this simple picture the sound speed $s$ can be basically calculated in two ways. First, from the analysis of the collective excitations, as the phase speed at $\kv \rightarrow 0$ of the acoustic mode $\omega(k)$, whose dispersion relation, in turn, is derivable from a kinetic equation or from some other nonlocal formalism via the response function technique. Second, alternatively, the sound speed may be obtained from the fundamental thermodynamic relationship $s_{\textrm{therm}} = (\partial P/\partial n)_S$ where $P$ is the pressure and the subscript indicates constant entropy. It is not a priori required that the two approaches have to lead to the same exact result. In fact, the thermodynamic approach is predicated upon the assumption of a local thermodynamic equilibrium (LTE), with a locally definable temperature and isotropic pressure with a concomitant equation of state; no such assumption is inherent in the collective mode description. Furthermore, in the thermodynamic approach one also has to observe the difference between the adiabatic and isothermal sound speeds; no such distinction exists in the kinetic formulation. This dichotomy was at the root of the debate in the early days of plasma physics concerning the coefficient of the $k^2$ term in the plasmon dispersion relation \cite{vanKampen1957,Pines1952}. Here, in the weakly and strongly coupled regimes we will rely mostly on the less restrictive collective mode approach. However, in the absence of an approach justifiable for all domains, in the medium coupling range the thermodynamic formula will be invoked. An additional technique that avoids explicit reference either to detailed dynamics, or to the LTE, exploits exact sum rule relationships will also be presented.

\subsection{Collisionless kinetic approach}
Starting with the weakly coupled $\Gamma \ll 1$ regime, one is inclined to adopt the conventional RPA (Vlasov) formalism, changing only the interaction potential from Coulomb to Yukawa. In the ``cold plasma" ($T\rightarrow 0$) limit this gives the simple expression for the dielectric function
\begin{equation}
    \epsilon(\kv,\omega) = 1 - \frac{\omega_0^2}{\omega^2}\frac{\bar k^2}{\bar k^2 + \kappa^2}, \quad \bar k = ka
\label{eq:RPAcold}
\end{equation}
leading to the RPA sound speed $c_0$. Similarly, one expects that finite temperature effects will be correctly rendered by replacing eq.~\eqref{eq:RPAcold} by the finite temperature RPA expression
\begin{equation}
    \epsilon( \kv,\omega) = 1 - \frac{\omega_0^2a^2}{\bar k^2 + \kappa^2 }\int d^3 v   \frac{1 }{\mathbf {k\cdot v} - \omega } \mathbf k \cdot\frac{\partial f(\mathbf v) }{\partial \mathbf v} ,
    \label{eq:RPAwarm}
\end{equation}
which leads to the dispersion equation
\begin{equation}
    \frac{\bar k^2 + \kappa^2}{3\Gamma} = \frac{1}{\beta n} \chi(x).
    \label{eq:RPAdispersion}
\end{equation}
Here $\chi(x) = \chi'(x) + i\chi''(x)$ is the complex (``screened") Vlasov density response function,
\begin{equation}
    \frac{1}{\beta n} \chi(x) = - 1 + \frac{x}{\sqrt\pi} P\int_{-\infty}^{\infty}\, dt \frac{e^{-t^2}}{x - t} - i\sqrt{\pi}x e^{-x^2},
	\label{eq:chi0}
\end{equation}
with $x \in \mathbb{R}$ and
\begin{equation}
    x = \frac{\omega}{k}\frac{1}{\sqrt{2} v_{\textrm{th}}}.
    \label{eq:x_chi0}
\end{equation}
Separating the real and imaginary parts of eq.~\eqref{eq:chi0}, one may follow the standard perturbative approach, seeking first a solution for the real part of the frequency from the real part of eq.~\eqref{eq:RPAdispersion}. In contrast to the Coulomb case, where the ensuing derivation is formally independent of $\Gamma$, here the condition
\begin{equation}
    \Gamma \geq 1.17 \kappa^2
    \label{gkcondition1}
\end{equation} 
has to be satisfied for such a solution to exist in the long wavelength limit \cite{Golden2018}. This bound can be made more restrictive by imposing the requirement of small Landau damping. Then the allowed $\Gamma$, $\kappa$ combinations are restricted to the domain \cite{Golden2018}
\begin{equation}
    \Gamma > 2.03 \kappa^2 .
    \label{gkcondition2}
\end{equation}
Within this domain the sound speed can be expressed as
\begin{equation}
    s^2_{\textrm{RPA}} = \frac{c_0^2}{2} \left ( 1 + \sqrt{1 + \frac{12}{w^2}  } \right ),
    \label{eq:gksound}
\end{equation}  
showing that $s_{\textrm{RPA}} > c_0$ always. According to eqs.~\eqref{gkcondition2} and \eqref{wratio}, $w > 2.47$, and these results should be valid in Domain 2 of Fig.~\ref{domainsplot}. In addition, we note that in the $T \rightarrow 0 $ ($w\rightarrow \infty$) limit eq.~\eqref{eq:gksound} reduces to 
\begin{equation}
    s^2_{\textrm{RPA}} (w \rightarrow \infty) = c_0^2 + 3 v_{\textrm{th}}^2,
    \label{eq:gksound2}
\end{equation}
where the factor $3$ in front of $v_{\textrm{th}}$ is the same as that found in a Coulomb OCP dispersion relation \cite{vanKampen1957}, a result that will be commented on below.

In order to see the behavior of the system in the RPA in more detail and to make contact with the language of the MD simulations we have calculated $S(\kv,\omega)$ from the Vlasov response function $\chi_0(k,\omega)$, eq.~\eqref{eq:chi0}, by means of the Fluctuation-Dissipation Theorem
\begin{equation}
    S(\kv,\omega) = - \frac{1}{\pi n \beta \omega} \frac{\chi_0''(\kv,\omega)}{ | \epsilon(\kv,\omega)|^2 }.
    \label{eq:FDT}
\end{equation}
In Fig.~\ref{RPAclrplotsk05} we show color maps and line plots of the RPA $S(\kv,\omega)$ for $\kappa = 0.5$ at three values of $\Gamma$ belonging to Domain 1 (panels a) and d)), Domain 2 (panels c) and f)), and on the boundary between the two Domains (panels b) and e)). The dashed lines in the color maps represent the linear dispersion $\omega = s_{\textrm{RPA}} k$ with $s_{\textrm{RPA}}$ given by eq.~\eqref{eq:gksound}. These plots illustrate well the strong thermal effects of Domain 1 where no acoustic peak is observed in $S(\kv,\omega)$, but only a wide ``shoulder". This is identified by the large shaded regions for $ka < 0.2$ in panel a). The $ka$ values shown in panels d) - f) are those accessible in our MD simulations. At the boundary between Domain 1 and 2 an acoustic peak starts to form, but it is immediately broadened at higher $ka$ values. Finally, in Domain 2, instead, the thermal broadening is reduced and an acoustic peak is visible for small and finite $ka$ values, $ka < 0.4$. In the color maps this is represented by the thin ellipses beneath the dashed line.
\begin{figure*}[htb]
    \centering
    \includegraphics[width = \linewidth]{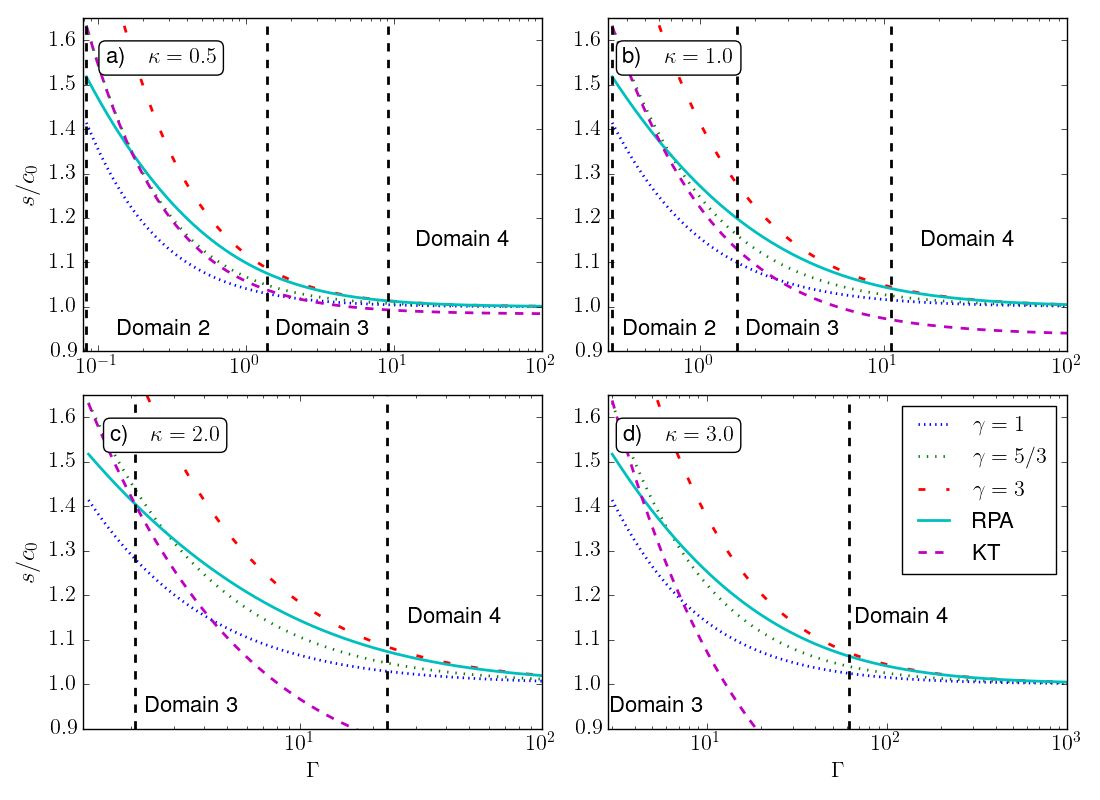}
    \caption{(Color online) Comparison between RPA sound \eqref{eq:gksound}, and fluid sound speeds \eqref{eq:hydrosound}, for an adiabatic $\gamma = 1$, isothermal $\gamma = 5/3$, process. $\gamma = 3$ corresponds to the sound speed from eq.~\eqref{eq:gksound2}. The KT line is the sound speed calculated in Ref.~\cite{Khrapak2015b,*Khrapak2016}. The dashed vertical lines represent the boundaries between the Domains. $\kappa$ values are indicated in the top left corner of each plot.}
    \label{fldvRPA}
\end{figure*}

\subsection{Collisional thermodynamic approach}
At this point one might ask whether there is, in fact, any good justification for using the RPA approximation for the weakly coupled Yukawa OCP. In contrast to the Coulomb case, where several proofs demonstrate that the RPA is the lowest order term in a formal expansion in terms of the coupling parameter \cite{Rostoker1960}, no such relationship has been established for Yukawa systems. Moreover, while the RPA neglects binary interactions between particles and assumes that the dynamics of the system is governed by the interaction via the average field only, in Yukawa systems the short range character of the interaction makes collisions more significant. This suggests that  the more phenomenological hydrodynamic description, which is predicated on the existence of a LTE maintained by frequent  collisions may represent an appropriate alternative.
Adopting this approach, one  is led to using the Euler equation combined with the continuity equation and the Helmholtz equation for the Yukawa potential.
We then obtain a sound speed of the form \cite{Khrapak2015b,*Khrapak2016}
\begin{equation}
    s^2_{\textrm{fluid}} = c_0^2 + \gamma L v_{\textrm{th}}^2 = c_0^2 \left ( 1 + \frac{\gamma L}{w^2} \right ),
    \label{eq:hydrosound}
\end{equation}
where the pressure term has been expressed in terms of the ratio of the specific heats, $\gamma = c_p/c_v$, and the inverse isothermal compressibility, $L = \beta (\partial P/\partial n)_\beta$.
To make any further progress an EOS is needed for the calculation of $\gamma$ and $L$. In general, $L$ consists of the ideal gas and correlational contributions, $L= 1 + L_c$, where $L_c$ is negative \cite{Rosenberg1997,Hartmann2005}. Here, in the weakly coupled regime we expect the perfect gas approximation to be valid, with $L = 1$. The value of $\gamma$ depends on whether an isothermal or adiabatic process is contemplated, corresponding to $\gamma = 1$ or $\gamma = 5/3$, respectively. Direct comparison between eq.~\eqref{eq:gksound} and eq.~\eqref{eq:hydrosound} indicates that the RPA sound speed exceeds the hydrodynamic sound speed for $\Gamma > 0.69\bar 4\kappa^2$. However, eq.~\eqref{eq:gksound} assumes \eqref{gkcondition1}--\eqref{gkcondition2}, therefore, the RPA sound speed will always be larger than the hydrodynamic sound in its Domain of validity. The concept of adiabatic process can also be generalized on phenomenological grounds in order to describe the system when LTE between longitudinal and transverse degrees of freedom on the time scale of wave propagation does not prevail. In this case it is the effective dimensionality associated with the wave propagation that determines the value of $\gamma$: $\gamma = 1 + 2/d$ ($d = $ dimensionality). Comparison with \eqref{eq:gksound2} shows that in Domain 2 a similar approach would yield $\gamma = 3$, \textit{i.e.} effective $d=1$ dimensionality. Such a situation is well known in Coulomb plasmas and may be considered the hallmark of LTE \cite{vanKampen1957}. 
Alternatively, eq.~\eqref{eq:hydrosound} is derivable from the fundamental thermodynamic relation $s_{\textrm{therm}} = (\partial P/\partial n )_S$, provided the pressure includes the effect of Yukawa field (to lowest order in the coupling). In this case, instead of using the field equation one has to add the Hartree (average field) contribution \cite{Rosenberg1997, Hartmann2005}
\begin{equation}
    P_{H} = \frac12 \frac{4\pi n^2q^2 a^2}{\kappa^2}
    \label{eq:HartreePressure}
\end{equation}
to the equation of state, yielding the additional Hartree contribution to $L$, providing $L = 1 + L_H$ and
\begin{equation}
    \frac{L_H}{\beta m} =\frac{\omega_0^2}{\kappa^2}
\end{equation}
which then leads back to eq.~\eqref{eq:hydrosound}. 

In the higher coupling regime, Domain 3, where the system is in the liquid state the RPA, which neglects correlations, is expected to overestimate the sound speed. As to the hydrodynamic approach, which also neglects correlations when $L = 1$, there is no indication on theoretical grounds what value of $\gamma$ to use in this Domain. One would expect that a value between the kinetic one-dimensional and the thermodynamic three-dimensional values, \textit{i.e.} $5/3 < \gamma < 3$, would be a reasonable choice. The underlying philosophy of LTE that may justify the use of hydrodynamic description in the weak coupling regime, where the thermal velocity dominates, may be invoked in the case of medium or strong coupling as well, although the onset of quasi-localization (see below) makes that assumption doubtful. Setting this point aside, the extension of the thermodynamic approach to the stronger coupling domain requires an EOS, valid for arbitrary $\kappa$ and $\Gamma$.
While a derivation from first principles is not available, various semi-phenomenological efforts \cite{Rosenfeld1998,*Rosenfeld2000,Khrapak2015a} have proved to be quite successful. Rosenfeld and Tarazona showed that the internal energy of systems with soft repulsive potential, including the Yukawa potential, exhibits a quasi--universal behavior \cite{Rosenfeld1998,*Rosenfeld2000}. The latter authors obtained a power law formula for the internal energy from fits to computer simulations. Khrapak and Thomas used this formula for the calculation of $\gamma$ and $L$ for a large range of $\Gamma$. Fig.~1 in Ref.~\cite{Khrapak2015b,*Khrapak2016} shows exactly that $5/3 < \gamma < 3 $ in the weakly to moderately coupled regime and $\gamma = 1$ in the strongly coupled regime. This observation seems to corroborate the assertion that the quasi-localized charge environment reduces the effective collision frequency and is not conducive to the formation of LTE.

A comparison between the RPA sound, eq.~\eqref{eq:gksound}, valid for $w > 2.47$, and the hydrodynamic sound, eq.~\eqref{eq:hydrosound} with $L = 1$ and different values of $\gamma$, is shown in Fig.~\ref{fldvRPA}. The sound speed calculated by Khrapak and Thomas \cite{Khrapak2015b,*Khrapak2016} is also plotted in Fig.~\ref{fldvRPA} as a dense dashed line (KT line). From these plots we note that the KT line overlaps with the adiabatic sound ($\gamma = 5/3$) in the weakly coupled regime while falling below the isothermal sound and the RPA value $c_0$ as $\Gamma$ increases. Furthermore, we note that the RPA sound approaches the $\gamma = 3$ line as $\kappa \rightarrow 0$ as given by eq.~\eqref{eq:gksound2}.  

\subsection{Sum rule based approach}
\begin{figure}[ht]
    \centering
    \includegraphics[width = \linewidth]{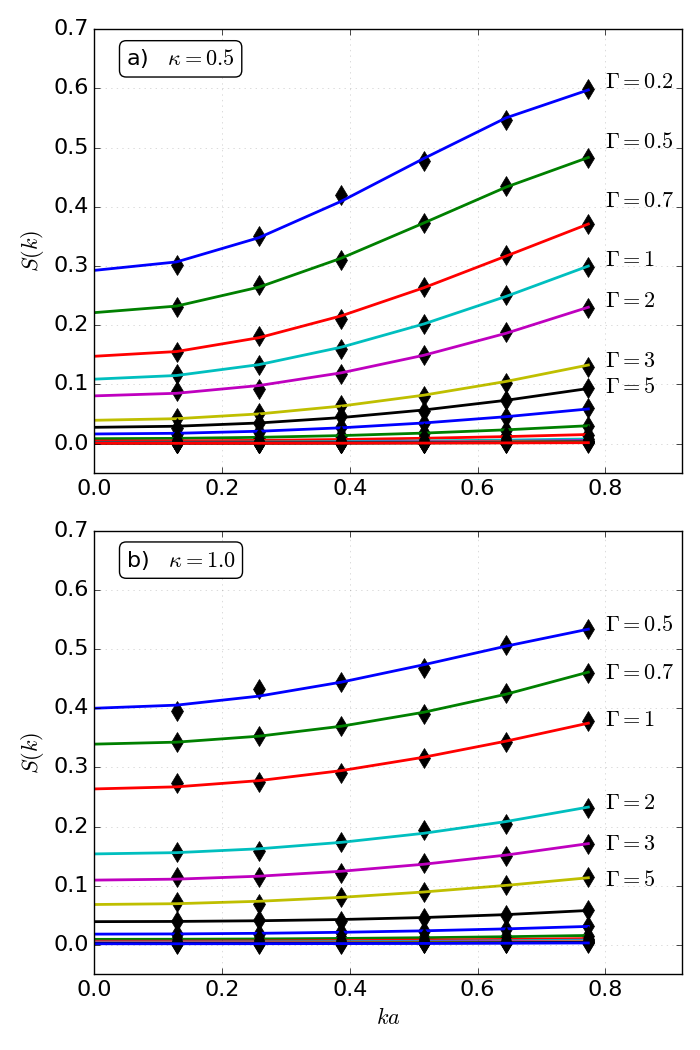}
    \caption{(Color online) Comparison between quartic fit (solid lines) and MD simulations (diamonds) for two $\kappa$ values. The values of $\kappa$ are indicated in the top left corner of each plot. For clarity only the first few $\Gamma$ values are indicated as the lines become densely packed at higher $\Gamma$. The missing labels are: $\Gamma = \{  10, 20, 40, 60, 80, 100, 120, 150, 200\}$  }
\label{Sk_Fits}
\end{figure}
An alternative approach, that still relies on the thermodynamic relationship, yields the sound speed in terms of  $L_T$, the (total) inverse isothermal compressibility \cite{Salin2007,Mithen2011b,HansenBook}
\begin{equation}
    s_F^2 = \frac{\gamma L_T}{\beta m },
\end{equation}
with $\gamma = 1$. Explicit reference to any EOS, here, can be avoided by invoking the static structure factor, $S(\kv)$, obtainable directly from MD simulations, and making use of the compressibility sum rule 
\begin{equation}
    \beta n L_T = \lim_{k \rightarrow 0} S(\kv),
\end{equation}
to find
\begin{equation}
    s^2_F = \lim_{k \rightarrow 0} \frac{(\beta m)^{-1}}{S(\kv)} =  \lim_{k \rightarrow 0} \frac{\omega_0^2 a^2}{3\Gamma} \frac{1}{S(\kv)}.
\label{eq:feynmansound}
\end{equation}
Remarkably, this very same relationship can be derived through the classical version \cite{Golden2003} of the Feynman Ansatz \cite{Feynman1954,*Feynman1956,*Cohen1957} as well, even though the main physical assumption in that model is quite different from the one implied by the hydrodynamic model. In the Feynman Ansatz no reference is made to LTE, rather it is assumed that the only significant contribution to the DSF comes from the collective excitation, in this case the sound wave $\omega_s(k) = s_F k$. In this case, the DSF reads
\begin{equation}
   S_{FA}(\kv,\omega) = A(\kv) \left [ \delta \left ( \omega - \omega(k) \right )+\delta \left ( \omega + \omega(k) \right ) \right ].
\end{equation}
Evoking, then, the compressibility sum rule
\begin{equation}
    \braket{\omega^{0}} = \int_{-\infty}^{\infty} d\omega S_{FA}(\kv,\omega) = S(\kv) \nonumber 
\end{equation}
\begin{equation}
	\implies A(k) = S(\kv),
\end{equation}
and the $f$-sum rule
\begin{equation}
    \braket{\omega^2} = \int_{-\infty}^{\infty} d\omega \omega^2 S_{FA}(\kv,\omega) = \frac{\omega_0^2 \bar{k}^2}{3\Gamma} \nonumber
\end{equation}
\begin{equation}
	\implies \omega^2_s(k) = \frac{\omega_0^2 \bar k^2}{3\Gamma S(\kv)} 
\end{equation}
one indeed recovers \eqref{eq:feynmansound} \cite{Murillo2010,Diaw2015}.

The  expression for the sound velocity given by \eqref{eq:feynmansound}, which will be referred to as the Feynman sound speed, is formally valid for all $\Gamma, \kappa$ values: all the information is embodied in $S(\kv)$. To obtain $S(\kv)$, three different methods have been implemented: \textit{(i)}  direct calculation of $\braket{n_{-\mathbf k}n_{\mathbf k} }$ (see Refs.~\cite{Hansen1975,Donko2008} for details), \textit{(ii)}  integration over the entire frequency range of $S(\kv,\omega)$ \textit{i.e.}
\begin{equation}
	S(\kv) = 2 \int_0^\infty d\omega S(\kv,\omega),
\end{equation}
\textit{(iii)} Fourier transformation  of the pair correlation function $h(r)$ \textit{i.e.}
\begin{equation}
    S(\kv) = 1 + n\int_0^\infty dr r^2 \frac{\sin(kr)}{kr}h(r).
\end{equation}
Each method has its drawbacks due to the finite integration ranges. No substantial differences, however, have been observed between the results of the first two methods, while those from the last have been found to be very poor at low $ka$. The results presented in Sec.~\ref{sec:results} are obtained using method \textit{(ii)}. The limit $k\rightarrow 0$  of $S(\kv)$, instead, is extrapolated by fitting a function $f_2(x) = a_0 + a_2x^2 + a_4x^4$ to the lowest few $ka$ values of the simulations. In Fig.~\ref{Sk_Fits} we show plots comparing MD data of $S(\kv)$ (diamonds) and fits to these data (solid lines) for two $\kappa$ values and the entire $\Gamma$ range. It is interesting to note the larger gap between the $\Gamma = 2$ and $\Gamma = 3$ lines in both plots, coinciding with the boundary between Domain 2 and Domain 3. This is more evident in the $\kappa = 1.0$ (third and fourth lines) plot than in $\kappa = 0.5$ (fifth and sixth lines). 

Before continuing it is important to clarify a point raised in a recent discussion on the compressibility \cite{Goree2014}. It is essential to realize the difference between the quantities $L_c$, $L = 1 + L_c$ and $L_T = 1 + L_c + L_H$, especially in view of a recent controversy on the sign of the compressibility \cite{Goree2014}, prompted perhaps by a somewhat loose usage of the term ``compressibility" in the literature. Note that $L_T$  is a positive quantity and must be so, both in order to guarantee the thermodynamic stability of the system and because of the compressibility sum rule that relates it to the positive definite $S(\kv)$. On the other hand, $L_c$ is always negative, while $L$ changes sign in the vicinity of the Kirkwood line as $\Gamma$ enters Domain 3. It is this latter effect that is usually referred to as the ``softening" of the EOS and claimed to be responsible for the ``lowering" of the sound speed with increasing $\Gamma$. In fact, the EOS never softens with higher density: it rather gets stiffer because of the dominance of the Hartree term. Similarly, the sound speed never decreases, only in relationship to $c_0$, which itself is a fast increasing function of the density. Furthermore, we note that the total pressure of the YOCP is always a positive quantity as well, due to presence of the Hartree term, eq.~\eqref{eq:HartreePressure}. This is at variance with the case of the Coulomb OCP, where only the contribution of the neutralizing background prevents the pressure from becoming negative at strong coupling. See also discussion in Ref.~\cite{Salin2007} and eq.~(6) in Ref.~\cite{Hartmann2005}. 

\subsection{Strong Coupling Theory: QLCA}
Turning now to the strongly coupled regime,  Domain 4, here correlations dominate and thermal motion plays a negligible role. An appropriate approach in this Domain is the Quasi-Localized Charge Approximation (QLCA) \cite{QLCA1990,*QLCA2000,*QLCA2000Err}. This technique has been quite successful in describing the collective excitations of charged particle systems \cite{Rosenberg1997,KalmanDeWitt2000,Rosenberg2016}. The basis of the formal development of the QLCA is the realization that the dominating feature of the strongly coupled state of a plasma is the quasi--localization of the charges. This, physical picture suggests a microscopic equation-of-motion model where the particles are trapped in local potential fluctuations. The particles occupy randomly located (but certainly not uncorrelated) sites and undergo oscillations around them. At the same time, however, the site positions also change and a continuous rearrangement of the underlying quasiequilibrium configuration takes place. Inherent in the QLC model is the assumption that the two time scales are well separated and that for the description of the fast oscillating motion, the time average (converted into ensemble average) of the drifting quasi-equilibrium configuration is sufficient. The application of the QLCA to the case of a YOCP yields a dynamical matrix $\mathbf C(\mathbf k)$ whose longitudinal component reads
\begin{equation}
    C(k) = \omega^2_0 \left \{ \frac{\bar k^2}{\bar k^2 + \kappa^2} + \int\, \frac{d\mathbf r}{4\pi q^2} \left ( \frac{ \bar{ \mathbf k }\cdot \nabla}{\bar k} \right )^2 \phi_r(r) h(r)  \right \}
    \label{QLCA_C}
\end{equation}
where $h(r)$ is the equilibrium pair correlation function \footnote{the factor $4\pi q^2$ in the denominator of the second term is due to our definition of $\phi(r) \propto q^2$ }. The collective mode is the solution to the equation $\omega^2 - C(k) = 0$ from which we obtain, after taking the limit $k\rightarrow 0$, the sound speed
\begin{equation}
    s_{\textrm{QLCA} } = c_0 \left \{ 1 + \frac{2\kappa^2}{15} \int d\bar r \mathcal K(\kappa \bar r)h(\bar r) e^{-\kappa \bar r} \right \}^{1/2},
\label{eq:QLCAsound}
\end{equation} 
\begin{equation}
    \mathcal K(\kappa \bar r) = \left [ 1 + \kappa \bar r + \frac34 (\kappa \bar r)^2 \right ],
\end{equation}
where $\bar r = r/a$. We emphasize that the correlational dependence comes only from the (negative) integral while the first term is the cold RPA sound. We observe that the QLCA result shows a monotonic decrease of the relative sound speed, as interparticle correlations increase -- a clear indication of the correlation induced ``softening" of the  equation of state (cf. discussion above). In the literature one finds different versions of the above equation. For example, the sound speed calculated from eq.~(29) of Ref.~\cite{Rosenberg1997} is obtained from the above eq.~\eqref{eq:QLCAsound} by rewriting the integral in terms of the correlational energy $E_c = (n/2) \int dr \phi(r) g(r)$ which in turn, leads again to the need of an EOS when $g(r)$ is not available. In Ref.~\cite{Rosenberg1997} the EOS used is the one calculated by Hamaguchi \textit{et al.} in Ref.~\cite{Hamaguchi1997}. In this work, we computed $g(r)$ directly from MD and therefore, we will use eq.~\eqref{eq:QLCAsound} for the calculation of the sound speed in the following. 

\subsection{Lattice}
Finally, arriving to Domain 5, the crystalline solid phase,  the theoretical description of the phonon excitations through the harmonic phonon approximation and lattice summation technique is well-known. It is believed that the lattice structure of a Yukawa solid is either BCC or FCC, depending on the value of the screening parameter $\kappa$. The appropriate phase diagram is due to Hamaguchi \textit{et al.} \cite{Hamaguchi1997}. According to this prediction a YOCP with $\kappa = 1$ crystallizes in a BCC, $\kappa = 3$ in an FCC lattice. The calculation of the corresponding phonon spectrum  is straightforward. The sound speed is given by 
\begin{equation}
s_{\rm solid} = \omega_L(k \rightarrow 0) / k
\end{equation}
where $\omega_L(k \rightarrow 0 )$ is the longitudinal mode of the dispersion relation 
\begin{equation}
|| \omega^2 \delta_{\mu\nu} - D_{\mu\nu}(\mathbf k) || = 0
\end{equation}
with the dynamical matrix
\begin{equation}
D_{\mu\nu} = - \frac 1m \sum_i \frac{\partial^2 \phi(r)}{\partial r_\mu \partial r_{\nu} } \left [ e^{-i\mathbf{k} \cdot \mathbf {r} } - 1\right ] 
\end{equation}

\begin{figure*}[ht]
    \centering
    \includegraphics[width = \linewidth]{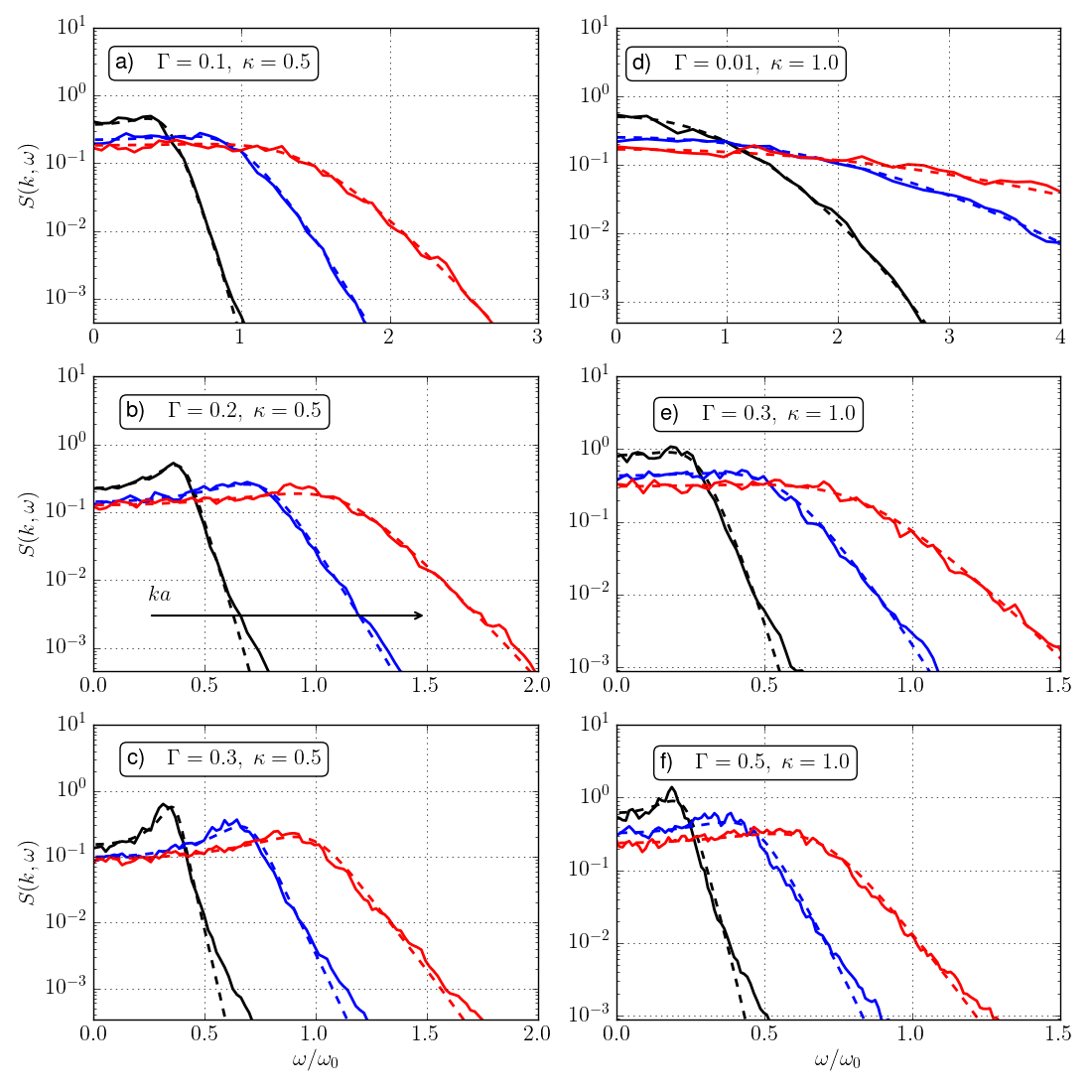}
    \caption{(Color online) Plots comparing $S(\kv,\omega)$ obtained from MD simulations (solid lines) and from RPA calculations eq.~\eqref{eq:FDT} (dashed lines), at the three lowest $ka$ values: $ka = \left \{ 0.1289, 0.2578, 0.3867 \right \}$. The arrow indicates direction of increasing $ka$. $\Gamma$ and $\kappa$ values are indicated in the top left corner of each plot.} 
    \label{fig:MDvRPA_Skw}
\end{figure*}

\section{Results}
\label{sec:results}
In this section we compare the previously described theoretical models with results from MD simulations; starting from a direct comparison between RPA calculations and MD simulations of $S(\kv,\omega)$, followed by a comparison of the sound speed across coupling regimes a similar investigation of the damping behavior.

\subsection{Dynamical Structure Function}
In the weakly coupled regime, the analytic formula \eqref{eq:FDT} allows for a direct comparison between the RPA calculations and MD simulations. This is shown in Fig.~\ref{fig:MDvRPA_Skw} where we present plots of $S(\kv,\omega)$ at six $\Gamma,\kappa$ points in parameter space. Dashed lines represent RPA calculations while solid lines MD simulations. The three left panels a) -- c) in the Figure represent $\Gamma,\kappa$ points belonging to Domain 2, while the three right panels indicates points in Domain 1 ( panel d) ), Domain 2 ( panel e) ), and on the boundary between the two Domains ( panel f) ). In all six plots the RPA faithfully reproduces the profile of $S(\kv,\omega)$ over a large range of frequencies with the only differences being in the bottom panels where the RPA shows a faster decay than the MD. Furthermore, the plots indicate the absence of a well defined acoustic peak in Domain 1 and on the boundary with Domain 2 (top panels and center right panel). Only when the system is well into Domain 2 a well defined acoustic peak  observed. MD simulations at $\Gamma$ values higher than those in the Figure, while still belonging to Domain 2, start to show deviation from the RPA formula indicating the onset of correlational effects as shown in Fig.~\ref{fig:MDDamping}.

\subsection{Sound Speed} 
\begin{figure*}[ht]
    \centering
    \includegraphics[width = .9\linewidth]{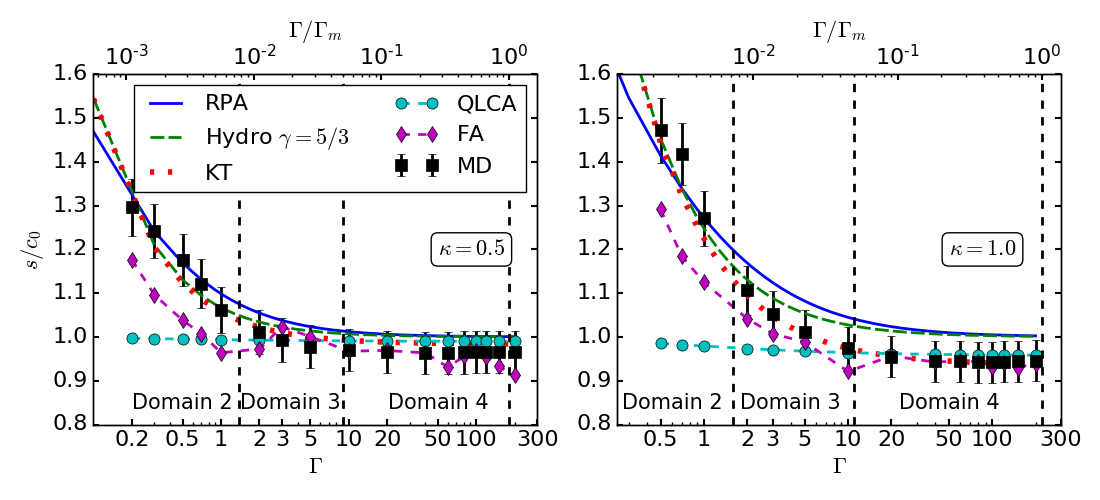}
    \caption{(Color online) Comparison between theoretical models and MD simulations (black squares) with 5\% errorbars. Blue solid line (RPA) corresponds to eq.~\eqref{eq:gksound}, green dashed line (Hydro) to \eqref{eq:hydrosound}, red dotted line (KT) to Ref.~\cite{Khrapak2015b,*Khrapak2016}, cyan circles (QLCA) to eq.~\eqref{eq:QLCAsound}, magenta diamonds (Feynman Ansatz) to eq.~\eqref{eq:feynmansound}. $\kappa$ values are indicated in the inserts.}
    \label{fig:results2}
\end{figure*}

Plots of the relative sound speed across three domains are presented in Fig.~\ref{fig:results2}, where the vertical error bars indicate a 5\% error for the MD sound speed. Five different theoretical models are presented: \textit{(i)} The line marked ``RPA'' is calculated from eq.~\eqref{eq:gksound}; \textit{(ii)} the line ``Hydro" corresponds to the fluid model, eq.~\eqref{eq:hydrosound}, taken at weak coupling and strict 3D thermodynamic equilibrium; \textit{(iii)} the line ``KT'' corresponds to the sound speed calculated from the EOS proposed by Khrapak and Thomas in Ref.~\cite{Khrapak2015a} for all $\Gamma$ values, with $\gamma$ determined from the EOS under the assumption of strict 3D thermodynamic equilibrium; \textit{(iv)} the ``QLCA'' line is calculated from eq.~\eqref{eq:QLCAsound}; and \textit{(v)} the ``FA" line represents the Feynman formula with the understanding of a prevailing isothermal condition.

\begin{figure*}[ht]
    \centering
    \includegraphics[width = .9\linewidth]{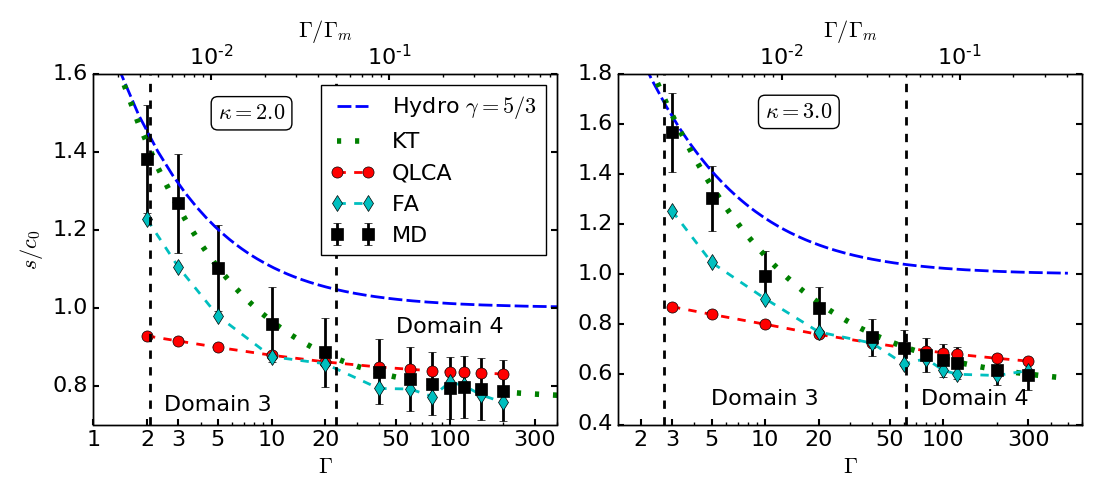}
    \caption{(Color online) Same as Fig~\ref{fig:results2}, but for $\kappa = 2.0, 3.0$. In this plot errorbars indicate a 10\% error.}
    \label{fig:results5}
\end{figure*}

As shown above, no acoustic peak is found in simulations pertaining to Domain 1 and only when the system enters Domain 2 does a well defined acoustic peak become visible. Fig.~\ref{fig:results2} further validates the plots in Fig.~\ref{fig:MDvRPA_Skw}. The RPA sound speed is found to be in very good agreement with the MD data for $\Gamma$ values close to the lower boundary of Domain 2, while it overestimates, despite remaining within the error bars, for higher $\Gamma$ values. Given the good agreement with the RPA sound, the slight underestimate of the hydrodynamic sound is expected, since, we recall that for $\Gamma = 0.69\bar4 \kappa^2$, the sound speed given by eq.~\eqref{eq:gksound} is always higher than the one calculated from eq.~\eqref{eq:hydrosound}. The Feynman sound, which is obtained directly from MD data without any reference to any particular EOS and thus should reproduce the MD results well, is, in fact, an underestimate, probably because of the assumed isothermal behavior. For $\Gamma$ values beyond Domain 2, both the RPA and the hydrodynamic formula deviate from MD results and overestimate the sound speed, as expected, since both of them ignore correlations. In contrast, in this Domain the Feynman sound falls within the 5\% error. The phenomenological formula of Khrapak and Thomas \cite{Khrapak2015b,*Khrapak2016} is in very good agreement with MD over the moderately and strongly coupled regimes. The underestimate in the weakly coupled regime is again expected given that their EOS approaches the ideal EOS at low $\Gamma$. Note that the intersection with the $=1$ line corresponds to the onset of negative compressibility: for the lowest $\kappa$ value KT seems to overestimate the $\Gamma$ value, where this happens, but it is in good agreement with MD data for higher $\kappa$ values. As for the QLCA speed, eq.~\eqref{eq:QLCAsound}, we find very good agreement only in Domain 4 since the quasi-localization model is not appropriate in Domains 2 and 3. Reaching the boundary of Domain 5, the KT and Feynman join smoothly the lattice value of the sound speed, but the QLCA seems to end up at a higher speed. Since the transition through the phase boundary should be smooth, at least in the case of the BCC, this requires some explanation. Observing that the discrepancy is most pronounced for small $\kappa$, where the linear portion of the dispersion curve is the shortest, we suggest that the reason for the mismatch may be that the smallest $ka$ value reached by simulations, $ka_{\textrm {min} } = 0.1289$, is not small enough to fall within the linear dispersion portion of the mode and therefore it provides an average slope, which is lower than the acoustic one. 

Comparison at larger $\kappa$ for Domains 3 and 4 are shown in Fig.~\ref{fig:results5}. Using the upper scale $\Gamma/\Gamma_m$ in the panels to find corresponding points, not much $\kappa$-dependence can be discerned in Domain 3. In Domain 4, though, especially near the freezing boundary, the sound speed decreases sharply as a function of $\kappa$. A similar effect has been reported already in Ref.~\cite{Hamaguchi1997,Rosenberg1997,Hartmann2005}.

\begin{figure}[ht]
    \centering
    \includegraphics[width = \linewidth]{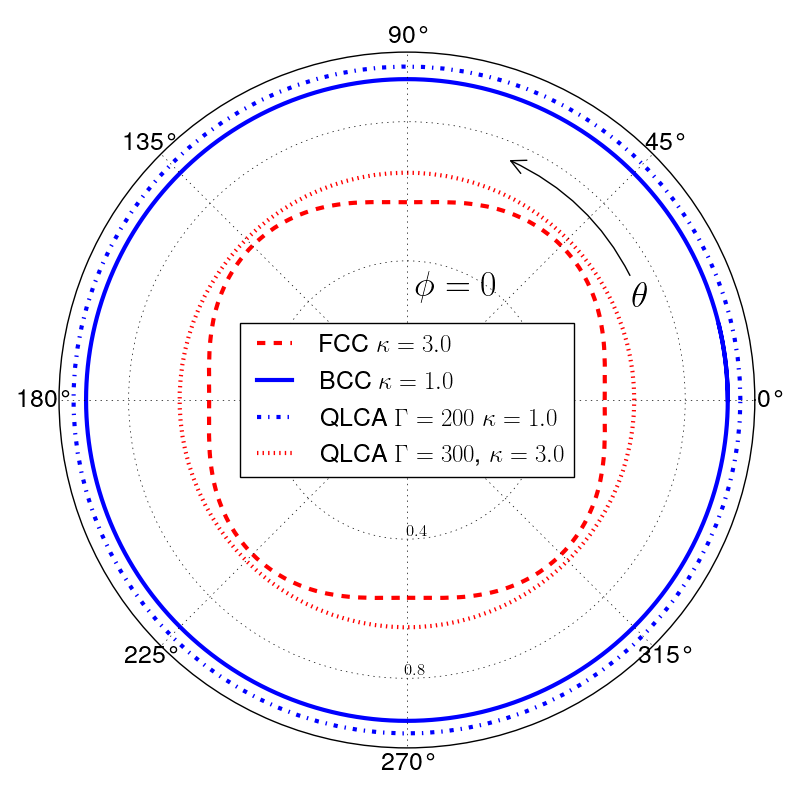}
    \caption{(Color online) Polar Plot comparing the lattice sound speeds and the QLCA speeds. $\theta$ represents the polar and $\phi$ the azimuthal angles. Inner lines refer to an FCC and outer lines to a BCC crystalline structure.} 
    \label{fig:LatticeFig}
\end{figure}

Finally in Fig.~\ref{fig:LatticeFig} we show a polar plot comparing the lattice sound speeds for two different lattice structures and the QLCA speed at the highest $\Gamma$ values, near the freezing point. The main difference between the BCC and FCC behavior is that in the former the sound speed is isotropic, while in the latter it is not, although the degree of anisotropy is small ($\sim$ 10\%). At the freezing boundary the QLCA predictions and lattice calculation results are in good agreement (within 10\%), (except for $\kappa=0.5$, not shown here, but discussed above); there is certainly no abrupt  change in the sound speed at the phase transition boundary, although obviously no exact match between the isotropic QLCA and the anisotropic FCC can exist.

\subsection{Damping}
\label{subsec:damping}
Determining the damping of the sound wave excitation is a delicate matter, fraught with two serious issues. First, the notion of ``damping" of a collective mode is ill-defined. On the one hand, theoretically, the damping can be calculated from the imaginary part of the complex frequency solution $\nu$ of the dispersion equation, $\varepsilon(k,\tilde \omega) = 0$ with $\tilde\omega(k) = \omega_{\textrm r}(k) + i \nu(k)$. On the other hand, the damping can be inferred either from MD simulations, or from an actual experiment, by examining the width of the acoustic peak in the $S(\kv,\omega)$ spectrum. The two definitions must yield different results as it is evident from eq.~\eqref{eq:FDT}. The second problem derives from the fact that the theoretical understanding of the damping mechanism in the medium or strong coupling regimes is lacking. While collisions seem to be the principal agent responsible for the damping of oscillations, other processes, such as trapping and caging, localization, self-diffusion, and non-linear mode-mode interactions may play significant roles as well. Thus, in the absence of theoretical guidance, the organization of observational data becomes difficult.
\begin{figure}[ht]
    \centering
    \includegraphics[width = \linewidth]{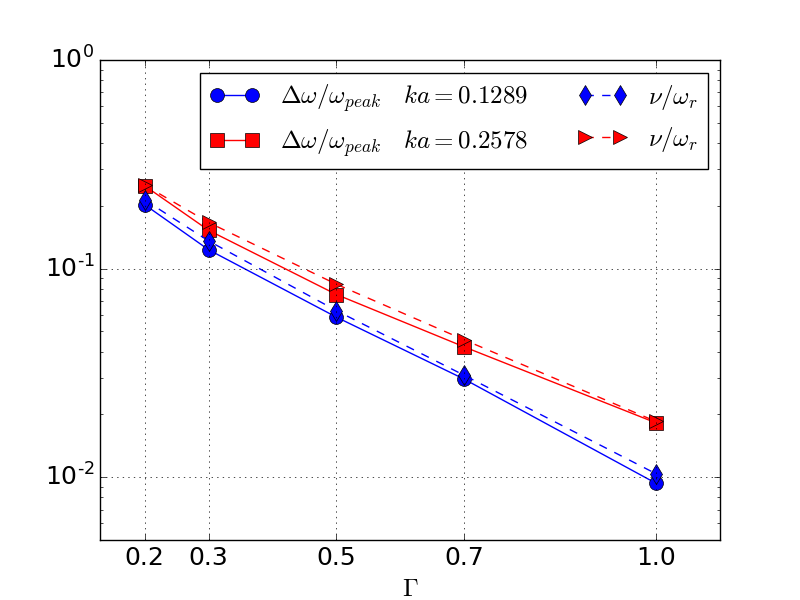}
    \caption{(Color online) Comparison between the relative (right) HWHM obtained from the RPA $S(\kv,\omega)$ and the ratio of the imaginary over the real part of the complex solution of $\varepsilon (k,\omega) = 0$. $\kappa = 0.5$. 	} 
    \label{fig:RPA_Damping_Comparison}
\end{figure}

Analytically, the shape of the ``spectral line" associated with  the sound wave is determined by the FDT, as given by eq.~\eqref{eq:FDT}. The shape generated by the complex pole due to the zero of the denominator would be a Lorentzian with the Half Width at Half Maximum (HWHM) set by $\nu$, but in fact the HWHM is also affected by the $\omega$--dependence of the numerator. The precise analytic form of this latter is known only for the RPA, where it mirrors the velocity distribution of the plasma particles. Evoking the analogy with spectral line formation, we may describe the effect as ``Doppler broadening" and characterize it with the parameter $\sigma$. Furthermore, in the weakly to moderately coupled regime, the acoustic peak in $S(\kv,\omega)$ is asymmetric with the right hand side decaying much faster than the left hand side (see Fig.~\ref{fig:MDvRPA_Skw}). Therefore, we have defined with HWHM the width $\Delta \omega = \omega_{\textrm{half}} - \omega_{\textrm{peak}}$ where the $\omega_{\textrm{half}}$ is the frequency at which the right hand side of the peak becomes half the amplitude of the peak.

In the following the damping is calculated in two different ways, depending on whether one is within the RPA, or outside the RPA region. For the RPA, the analytic form of $S(\kv,\omega)$ being given, the calculations of both $\nu$ and of the HWHM are straightforward. Outside the RPA, the damping is determined through fitting a Voigt distribution function, which is designed to combine Doppler broadening with intrinsic damping,
\begin{equation}
    V (x,A,\omega_k,\sigma,\delta) = \frac{A {\operatorname{Re} }\left [ \mathcal W (z) \right ]}{\sigma \sqrt{2\pi} }, \; z = \frac{x - \omega_k + i\delta}{\sqrt{2\sigma^2} },
\end{equation}
where $\mathcal W(z)$ is the Faddeeva function, $A$ the amplitude, and $\omega_k$ the peak location \cite{NIST:DLMF,*Fried1961,*Faddeyeva1961}. The parameter $\sigma$ indicates the effects of Doppler broadening due to the thermal distribution of particle velocities, while the parameter $\delta$ indicates the intrinsic damping of the mode. In the low coupling regime, \textit{i.e.} values close to lower boundary of Domain 3, the Voigt profile has been multiplied by an inverse power law $ \propto \omega^{-p}$, $p >0$, in order to help reproduce the asymmetry of the MD peak (see bottom panels in Fig.~\ref{fig:MDvRPA_Skw}). 

Earlier we have found that in Domain 1 the large damping quenches the acoustic mode. In Domain 2 the two measures of the damping, both $\nu$ and the HWHM, as calculated analytically from the RPA formulas \eqref{eq:RPAdispersion} and \eqref{eq:FDT}, are shown in Fig 10. There is very little difference between the two results. It should be remembered, though, that the strong asymmetry of line shape is not properly accounted for in the present calculational protocol and therefore the HWHM result should be regarded as an underestimate. The damping mechanism in this Domain is most likely Landau damping, since $\Gamma$ is small and $w > 1$. Furthermore, the faithful reproduction of the $S(\kv,\omega)$ profile by the RPA corroborates this assertion.

\begin{figure*}[ht]
    \centering
    \includegraphics[width = .9\textwidth]{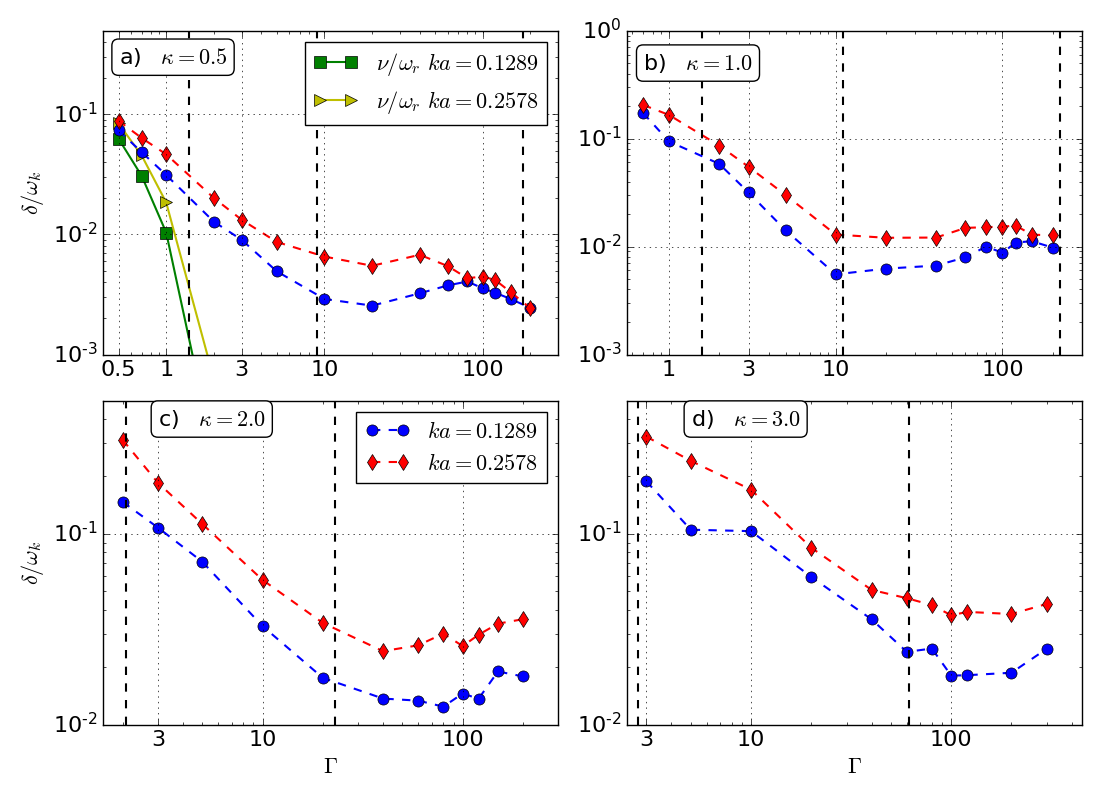}
    \caption{(Color online) Damping coefficient extracted from fits to MD simulations at the two lowest $ka$ values. The top left panel shows also the exponential decay of the (Landau) damping obtained from the complex solution of $\varepsilon(k,\omega) = 0$. $\kappa$ values are indicated in the top left corner of each plot.} 
    \label{fig:MDDamping}
\end{figure*}
\begin{figure*}[ht]
    \centering
    \includegraphics[width = .9\textwidth]{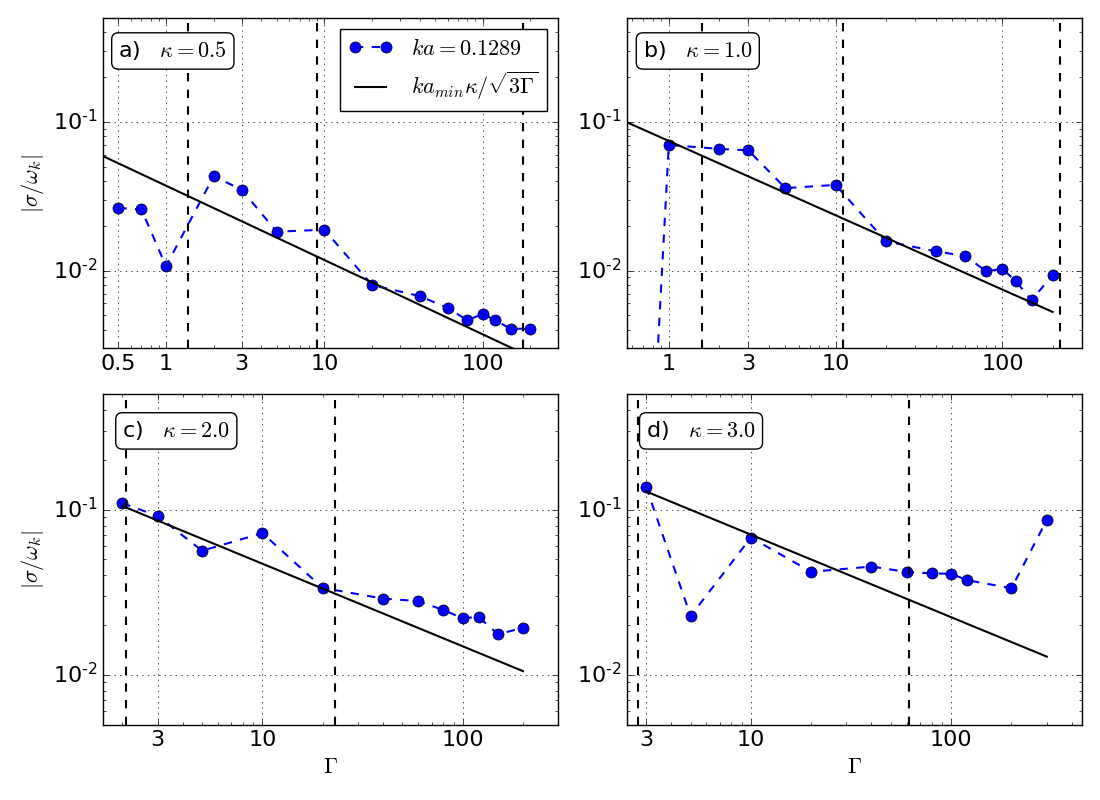}
    \caption{$\sigma$ parameter extracted from fits to MD simulations at the lowest $ka$ value.} 
    \label{fig:MDDamping_sigma}
\end{figure*} 

For higher $\Gamma$ values the ``intrinsic damping" $\delta$ and the ``Doppler width" $\sigma$ as obtained from the Voigt profile are shown in Fig.~\ref{fig:MDDamping} and Fig.~\ref{fig:MDDamping_sigma} respectively. As $\Gamma$ increases beyond the Domain 2/Domain 3 boundary Landau damping decreases exponentially with $\Gamma$ (from eq.~\eqref{eq:x_chi0} $x \propto w = \sqrt{3\Gamma}/\kappa$) and fails to account for the observed damping, see top left panel in Fig.~\ref{fig:MDDamping}. In Domain 2, near the boundary, and in Domain 3 correlations set on, with the ensuing expected increased collision frequency. Here, the hydrodynamic description leads to a better description of the sound speed and thus we expect collisions to be the dominant mechanism, not only for the creation, but also for the damping of the acoustic mode. Furthermore, in the moderately coupled regime, one would expect the collision frequency to increase with $\Gamma$ \cite{Chapman1990} with an ensuing growth of the damping.  This behavior has certainly been theoretically predicted for the COCP \cite{Oberman1962, DuBois1962, Coste1965b} and there is little doubt that the same kind of analysis would apply to the YOCP. However, in the current case we  fail to observe a region with such $\Gamma$--dependence.  Rather, a monotonic decay sets on right  after the Landau damping dominated RPA region. The decay of the damping with $\Gamma$ may be explained by a physical picture where with increasing correlations particles get more and more trapped in local potential minima and thereby collisions become less frequent. This process  may go on until the system enters the strongly coupled regime of Domain 4, at which point the completed quasi-localization of the particles settles the damping around a constant value. Nevertheless, the absence of an intermediate  region between Landau damping and quasi-localization where the damping grows with increasing $\Gamma$ is puzzling: substantially more work and simulations are needed to understand this behavior and its seeming disagreement with theoretical expectations.
 
\section{Concluding Remarks}
\label{sec:conclusion}
In this work we have investigated the sound speed of a YOCP over a wide range of $\Gamma$ (coupling) and $\kappa$ (screening) values. The former spans over states from the very weak, virtually ideal gas-like all the way to the extremely strong crystalline solid phases. We believe that this is the first time that, with the aid of advanced simulation technique and a combination of various theoretical models, such a comprehensive description of the evolution of a collective mode across coupling values have become possible. We have identified five Domains in the $\Gamma$-$\kappa$ parameter space. At weak coupling long wavelength density oscillations are possible only  with a phase speed $c_0$  greater than the thermal speed of particles $v_{\textrm{th}}$. In marked contrast to Coulomb systems, this requires a minimum value of $\Gamma$, below which the thermal motion quenches collective excitations. Once  $\Gamma$ exceeds this value, the YOCP can support a mode with an acoustic-like dispersion. The phase  speed of this mode, in general, is given by the combination  of three contributions: a coupling (temperature) independent mean field term $\omega_0a/\kappa$, a (negative) correlational, and a (positive) thermal coupling  dependent term. In the weak coupling regime the second term is absent, while the latter is derivable from a hydrodynamical description with an ideal gas equation of state. This indicates that the system behaves as a gas in local thermodynamic equilibrium sustained by frequent collisions. At the same time, at reasonably low values of the screening parameter the YOCP resembles a weakly coupled Coulomb gas with a dispersion relation derivable from the collisionless Vlasov kinetic equation.
As the coupling parameter is increased beyond the Kirkwood value $\Gamma_{\textrm K}$ (defined as the value at which the $r \rightarrow \infty$ behavior of the pair correlation function $h(r)$ becomes oscillatory) the system enters the moderately coupled liquid regime, in which particle correlations become of the same order  as the thermal effects.  Further increase of the coupling leads to the strongly coupled liquid domain, with the onset of quasi--localization. In this domain the thermal contribution to the sound speed becomes negligible.
Finally, for $\Gamma > \Gamma_m$ the system crystallizes with a lattice structure determined by the screening parameter. Depending on the lattice structure, the resulting  acoustic phonon shows an anisotropic (FCC) or isotropic (BCC) spectrum,  with a value in the very vicinity of the sound speed in the strongly coupled liquid.

Preliminary investigation of the damping, measured as the HWHM of the acoustic peak in $S(\kv,\omega)$, corroborates the physical picture where at low coupling in the gaseous state Landau type damping dominates, with a gradual enhancement of the collisional induced damping as $\Gamma$ increases. Contrary to naive expectation though, (based partly on the available exact perturbational results pertaining to Coulomb systems, which show an increase in damping with increasing $\Gamma$) the damping exhibits a monotonic decay all the way through the strongly coupled domain. This behavior may be attributed to the onset of quasi--localization, which is expected to reduce the collision frequency between particles isolated from each in distinct cages. The details of the coupling dependence of the damping are, however,  still not well understood.

Our work relies on the analysis of a huge body of MD data, which complement the work found in the literature inasmuch as our simulations extend well into the so far simulationaly unexplored weakly coupled $\Gamma < 1$ regime.
\begin{acknowledgements}
	This research was supported by Grant No. NSF PHY-1613102 and PHY-1740203, and by the Hungarian National Office for Research, Development, and Innovation, via Grants No. NKFIH K119357 and  No. NKFIH K115805.
\end{acknowledgements}


\begin{thebibliography}{51}%
\makeatletter
\providecommand \@ifxundefined [1]{%
 \@ifx{#1\undefined}
}%
\providecommand \@ifnum [1]{%
 \ifnum #1\expandafter \@firstoftwo
 \else \expandafter \@secondoftwo
 \fi
}%
\providecommand \@ifx [1]{%
 \ifx #1\expandafter \@firstoftwo
 \else \expandafter \@secondoftwo
 \fi
}%
\providecommand \natexlab [1]{#1}%
\providecommand \enquote  [1]{``#1''}%
\providecommand \bibnamefont  [1]{#1}%
\providecommand \bibfnamefont [1]{#1}%
\providecommand \citenamefont [1]{#1}%
\providecommand \href@noop [0]{\@secondoftwo}%
\providecommand \href [0]{\begingroup \@sanitize@url \@href}%
\providecommand \@href[1]{\@@startlink{#1}\@@href}%
\providecommand \@@href[1]{\endgroup#1\@@endlink}%
\providecommand \@sanitize@url [0]{\catcode `\\12\catcode `\$12\catcode
  `\&12\catcode `\#12\catcode `\^12\catcode `\_12\catcode `\%12\relax}%
\providecommand \@@startlink[1]{}%
\providecommand \@@endlink[0]{}%
\providecommand \url  [0]{\begingroup\@sanitize@url \@url }%
\providecommand \@url [1]{\endgroup\@href {#1}{\urlprefix }}%
\providecommand \urlprefix  [0]{URL }%
\providecommand \Eprint [0]{\href }%
\providecommand \doibase [0]{http://dx.doi.org/}%
\providecommand \selectlanguage [0]{\@gobble}%
\providecommand \bibinfo  [0]{\@secondoftwo}%
\providecommand \bibfield  [0]{\@secondoftwo}%
\providecommand \translation [1]{[#1]}%
\providecommand \BibitemOpen [0]{}%
\providecommand \bibitemStop [0]{}%
\providecommand \bibitemNoStop [0]{.\EOS\space}%
\providecommand \EOS [0]{\spacefactor3000\relax}%
\providecommand \BibitemShut  [1]{\csname bibitem#1\endcsname}%
\let\auto@bib@innerbib\@empty
\bibitem [{\citenamefont {Morfill}\ and\ \citenamefont
  {Ivlev}(2009)}]{Morfill2009}%
  \BibitemOpen
  \bibfield  {author} {\bibinfo {author} {\bibfnamefont {G.~E.}\ \bibnamefont
  {Morfill}}\ and\ \bibinfo {author} {\bibfnamefont {A.~V.}\ \bibnamefont
  {Ivlev}},\ }\href {\doibase 10.1103/RevModPhys.81.1353} {\bibfield  {journal}
  {\bibinfo  {journal} {Rev. Mod. Phys.}\ }\textbf {\bibinfo {volume} {81}},\
  \bibinfo {pages} {1353} (\bibinfo {year} {2009})}\BibitemShut {NoStop}%
\bibitem [{\citenamefont {Murillo}(2010)}]{Murillo2010}%
  \BibitemOpen
  \bibfield  {author} {\bibinfo {author} {\bibfnamefont {M.~S.}\ \bibnamefont
  {Murillo}},\ }\href {\doibase 10.1103/PhysRevE.81.036403} {\bibfield
  {journal} {\bibinfo  {journal} {Phys. Rev. E}\ }\textbf {\bibinfo {volume}
  {81}},\ \bibinfo {pages} {036403} (\bibinfo {year} {2010})}\BibitemShut
  {NoStop}%
\bibitem [{\citenamefont {Killian}\ \emph {et~al.}(2007)\citenamefont
  {Killian}, \citenamefont {Pattard}, \citenamefont {Pohl},\ and\ \citenamefont
  {Rost}}]{Killian2007}%
  \BibitemOpen
  \bibfield  {author} {\bibinfo {author} {\bibfnamefont {T.}~\bibnamefont
  {Killian}}, \bibinfo {author} {\bibfnamefont {T.}~\bibnamefont {Pattard}},
  \bibinfo {author} {\bibfnamefont {T.}~\bibnamefont {Pohl}}, \ and\ \bibinfo
  {author} {\bibfnamefont {J.}~\bibnamefont {Rost}},\ }\href {\doibase
  https://doi.org/10.1016/j.physrep.2007.04.007} {\bibfield  {journal}
  {\bibinfo  {journal} {Physics Reports}\ }\textbf {\bibinfo {volume} {449}},\
  \bibinfo {pages} {77 } (\bibinfo {year} {2007})}\BibitemShut {NoStop}%
\bibitem [{\citenamefont {Salin}(2007)}]{Salin2007}%
  \BibitemOpen
  \bibfield  {author} {\bibinfo {author} {\bibfnamefont {G.}~\bibnamefont
  {Salin}},\ }\href {\doibase 10.1063/1.2759881} {\bibfield  {journal}
  {\bibinfo  {journal} {Physics of Plasmas}\ }\textbf {\bibinfo {volume}
  {14}},\ \bibinfo {pages} {082316} (\bibinfo {year} {2007})},\ \Eprint
  {http://arxiv.org/abs/http://dx.doi.org/10.1063/1.2759881}
  {http://dx.doi.org/10.1063/1.2759881} \BibitemShut {NoStop}%
\bibitem [{\citenamefont {Mithen}\ \emph
  {et~al.}(2011{\natexlab{a}})\citenamefont {Mithen}, \citenamefont
  {Daligault},\ and\ \citenamefont {Gregori}}]{Mithen2011a}%
  \BibitemOpen
  \bibfield  {author} {\bibinfo {author} {\bibfnamefont {J.~P.}\ \bibnamefont
  {Mithen}}, \bibinfo {author} {\bibfnamefont {J.}~\bibnamefont {Daligault}}, \
  and\ \bibinfo {author} {\bibfnamefont {G.}~\bibnamefont {Gregori}},\ }\href
  {\doibase 10.1103/PhysRevE.83.015401} {\bibfield  {journal} {\bibinfo
  {journal} {Phys. Rev. E}\ }\textbf {\bibinfo {volume} {83}},\ \bibinfo
  {pages} {015401} (\bibinfo {year} {2011}{\natexlab{a}})}\BibitemShut
  {NoStop}%
\bibitem [{\citenamefont {Mithen}\ \emph
  {et~al.}(2011{\natexlab{b}})\citenamefont {Mithen}, \citenamefont
  {Daligault}, \citenamefont {Crowley},\ and\ \citenamefont
  {Gregori}}]{Mithen2011b}%
  \BibitemOpen
  \bibfield  {author} {\bibinfo {author} {\bibfnamefont {J.~P.}\ \bibnamefont
  {Mithen}}, \bibinfo {author} {\bibfnamefont {J.}~\bibnamefont {Daligault}},
  \bibinfo {author} {\bibfnamefont {B.~J.~B.}\ \bibnamefont {Crowley}}, \ and\
  \bibinfo {author} {\bibfnamefont {G.}~\bibnamefont {Gregori}},\ }\href
  {\doibase 10.1103/PhysRevE.84.046401} {\bibfield  {journal} {\bibinfo
  {journal} {Phys. Rev. E}\ }\textbf {\bibinfo {volume} {84}},\ \bibinfo
  {pages} {046401} (\bibinfo {year} {2011}{\natexlab{b}})}\BibitemShut
  {NoStop}%
\bibitem [{\citenamefont {Rosenberg}\ and\ \citenamefont
  {Kalman}(1997)}]{Rosenberg1997}%
  \BibitemOpen
  \bibfield  {author} {\bibinfo {author} {\bibfnamefont {M.}~\bibnamefont
  {Rosenberg}}\ and\ \bibinfo {author} {\bibfnamefont {G.}~\bibnamefont
  {Kalman}},\ }\href {\doibase 10.1103/PhysRevE.56.7166} {\bibfield  {journal}
  {\bibinfo  {journal} {Phys. Rev. E}\ }\textbf {\bibinfo {volume} {56}},\
  \bibinfo {pages} {7166} (\bibinfo {year} {1997})}\BibitemShut {NoStop}%
\bibitem [{\citenamefont {Kaw}\ and\ \citenamefont {Sen}(1998)}]{Kaw1998}%
  \BibitemOpen
  \bibfield  {author} {\bibinfo {author} {\bibfnamefont {P.~K.}\ \bibnamefont
  {Kaw}}\ and\ \bibinfo {author} {\bibfnamefont {A.}~\bibnamefont {Sen}},\
  }\href {\doibase 10.1063/1.873073} {\bibfield  {journal} {\bibinfo  {journal}
  {Physics of Plasmas}\ }\textbf {\bibinfo {volume} {5}},\ \bibinfo {pages}
  {3552} (\bibinfo {year} {1998})}\BibitemShut {NoStop}%
\bibitem [{\citenamefont {Kalman}\ \emph {et~al.}(2000)\citenamefont {Kalman},
  \citenamefont {Rosenberg},\ and\ \citenamefont {DeWitt}}]{KalmanDeWitt2000}%
  \BibitemOpen
  \bibfield  {author} {\bibinfo {author} {\bibfnamefont {G.}~\bibnamefont
  {Kalman}}, \bibinfo {author} {\bibfnamefont {M.}~\bibnamefont {Rosenberg}}, \
  and\ \bibinfo {author} {\bibfnamefont {H.~E.}\ \bibnamefont {DeWitt}},\
  }\href {\doibase 10.1103/PhysRevLett.84.6030} {\bibfield  {journal} {\bibinfo
   {journal} {Phys. Rev. Lett.}\ }\textbf {\bibinfo {volume} {84}},\ \bibinfo
  {pages} {6030} (\bibinfo {year} {2000})}\BibitemShut {NoStop}%
\bibitem [{\citenamefont {Ohta}\ and\ \citenamefont
  {Hamaguchi}(2000)}]{Ohta2000}%
  \BibitemOpen
  \bibfield  {author} {\bibinfo {author} {\bibfnamefont {H.}~\bibnamefont
  {Ohta}}\ and\ \bibinfo {author} {\bibfnamefont {S.}~\bibnamefont
  {Hamaguchi}},\ }\href {\doibase 10.1103/PhysRevLett.84.6026} {\bibfield
  {journal} {\bibinfo  {journal} {Phys. Rev. Lett.}\ }\textbf {\bibinfo
  {volume} {84}},\ \bibinfo {pages} {6026} (\bibinfo {year}
  {2000})}\BibitemShut {NoStop}%
\bibitem [{\citenamefont {Murillo}(2000)}]{Murillo2000}%
  \BibitemOpen
  \bibfield  {author} {\bibinfo {author} {\bibfnamefont {M.~S.}\ \bibnamefont
  {Murillo}},\ }\href {\doibase 10.1063/1.873779} {\bibfield  {journal}
  {\bibinfo  {journal} {Physics of Plasmas}\ }\textbf {\bibinfo {volume} {7}},\
  \bibinfo {pages} {33} (\bibinfo {year} {2000})}\BibitemShut {NoStop}%
\bibitem [{\citenamefont {Anderson}(1963)}]{Anderson1963}%
  \BibitemOpen
  \bibfield  {author} {\bibinfo {author} {\bibfnamefont {P.~W.}\ \bibnamefont
  {Anderson}},\ }\href {\doibase 10.1103/PhysRev.130.439} {\bibfield  {journal}
  {\bibinfo  {journal} {Phys. Rev.}\ }\textbf {\bibinfo {volume} {130}},\
  \bibinfo {pages} {439} (\bibinfo {year} {1963})}\BibitemShut {NoStop}%
\bibitem [{\citenamefont {Mithen}(2014)}]{Mithen2014}%
  \BibitemOpen
  \bibfield  {author} {\bibinfo {author} {\bibfnamefont {J.~P.}\ \bibnamefont
  {Mithen}},\ }\href {\doibase 10.1103/PhysRevE.89.013101} {\bibfield
  {journal} {\bibinfo  {journal} {Phys. Rev. E}\ }\textbf {\bibinfo {volume}
  {89}},\ \bibinfo {pages} {013101} (\bibinfo {year} {2014})}\BibitemShut
  {NoStop}%
\bibitem [{\citenamefont {Donk{\'{o}}}\ \emph {et~al.}(2008)\citenamefont
  {Donk{\'{o}}}, \citenamefont {Kalman},\ and\ \citenamefont
  {Hartmann}}]{Donko2008}%
  \BibitemOpen
  \bibfield  {author} {\bibinfo {author} {\bibfnamefont {Z.}~\bibnamefont
  {Donk{\'{o}}}}, \bibinfo {author} {\bibfnamefont {G.~J.}\ \bibnamefont
  {Kalman}}, \ and\ \bibinfo {author} {\bibfnamefont {P.}~\bibnamefont
  {Hartmann}},\ }\href {\doibase 10.1088/0953-8984/20/41/413101} {\bibfield
  {journal} {\bibinfo  {journal} {Journal of Physics: Condensed Matter}\
  }\textbf {\bibinfo {volume} {20}},\ \bibinfo {pages} {413101} (\bibinfo
  {year} {2008})}\BibitemShut {NoStop}%
\bibitem [{\citenamefont {Diaw}\ and\ \citenamefont
  {Murillo}(2015)}]{Diaw2015}%
  \BibitemOpen
  \bibfield  {author} {\bibinfo {author} {\bibfnamefont {A.}~\bibnamefont
  {Diaw}}\ and\ \bibinfo {author} {\bibfnamefont {M.~S.}\ \bibnamefont
  {Murillo}},\ }\href {\doibase 10.1103/PhysRevE.92.013107} {\bibfield
  {journal} {\bibinfo  {journal} {Phys. Rev. E}\ }\textbf {\bibinfo {volume}
  {92}},\ \bibinfo {pages} {013107} (\bibinfo {year} {2015})}\BibitemShut
  {NoStop}%
\bibitem [{\citenamefont {Rosenberg}\ \emph {et~al.}(2016)\citenamefont
  {Rosenberg}, \citenamefont {Kalman}, \citenamefont {Hartmann},\ and\
  \citenamefont {Donk\'o}}]{Rosenberg2016}%
  \BibitemOpen
  \bibfield  {author} {\bibinfo {author} {\bibfnamefont {M.}~\bibnamefont
  {Rosenberg}}, \bibinfo {author} {\bibfnamefont {G.~J.}\ \bibnamefont
  {Kalman}}, \bibinfo {author} {\bibfnamefont {P.}~\bibnamefont {Hartmann}}, \
  and\ \bibinfo {author} {\bibfnamefont {Z.}~\bibnamefont {Donk\'o}},\ }\href
  {\doibase 10.1103/PhysRevE.94.033203} {\bibfield  {journal} {\bibinfo
  {journal} {Phys. Rev. E}\ }\textbf {\bibinfo {volume} {94}},\ \bibinfo
  {pages} {033203} (\bibinfo {year} {2016})}\BibitemShut {NoStop}%
\bibitem [{\citenamefont {Khrapak}\ and\ \citenamefont
  {Thomas}(2015{\natexlab{a}})}]{Khrapak2015b}%
  \BibitemOpen
  \bibfield  {author} {\bibinfo {author} {\bibfnamefont {S.~A.}\ \bibnamefont
  {Khrapak}}\ and\ \bibinfo {author} {\bibfnamefont {H.~M.}\ \bibnamefont
  {Thomas}},\ }\href {\doibase 10.1103/PhysRevE.91.033110} {\bibfield
  {journal} {\bibinfo  {journal} {Phys. Rev. E}\ }\textbf {\bibinfo {volume}
  {91}},\ \bibinfo {pages} {033110} (\bibinfo {year}
  {2015}{\natexlab{a}})}\BibitemShut {NoStop}%
\bibitem [{\citenamefont {Khrapak}(2016)}]{Khrapak2016}%
  \BibitemOpen
  \bibfield  {author} {\bibinfo {author} {\bibfnamefont {S.~A.}\ \bibnamefont
  {Khrapak}},\ }\href {\doibase 10.1088/0741-3335/58/1/014022} {\bibfield
  {journal} {\bibinfo  {journal} {Plasma Physics and Controlled Fusion}\
  }\textbf {\bibinfo {volume} {58}},\ \bibinfo {pages} {014022} (\bibinfo
  {year} {2016})}\BibitemShut {NoStop}%
\bibitem [{\citenamefont {Hansen}\ \emph {et~al.}(1975)\citenamefont {Hansen},
  \citenamefont {McDonald},\ and\ \citenamefont {Pollock}}]{Hansen1975}%
  \BibitemOpen
  \bibfield  {author} {\bibinfo {author} {\bibfnamefont {J.~P.}\ \bibnamefont
  {Hansen}}, \bibinfo {author} {\bibfnamefont {I.~R.}\ \bibnamefont
  {McDonald}}, \ and\ \bibinfo {author} {\bibfnamefont {E.~L.}\ \bibnamefont
  {Pollock}},\ }\href {\doibase 10.1103/PhysRevA.11.1025} {\bibfield  {journal}
  {\bibinfo  {journal} {Phys. Rev. A}\ }\textbf {\bibinfo {volume} {11}},\
  \bibinfo {pages} {1025} (\bibinfo {year} {1975})}\BibitemShut {NoStop}%
\bibitem [{\citenamefont {Vaulina}\ \emph {et~al.}(2002)\citenamefont
  {Vaulina}, \citenamefont {Khrapak},\ and\ \citenamefont
  {Morfill}}]{Vaulina2002}%
  \BibitemOpen
  \bibfield  {author} {\bibinfo {author} {\bibfnamefont {O.}~\bibnamefont
  {Vaulina}}, \bibinfo {author} {\bibfnamefont {S.}~\bibnamefont {Khrapak}}, \
  and\ \bibinfo {author} {\bibfnamefont {G.}~\bibnamefont {Morfill}},\ }\href
  {\doibase 10.1103/PhysRevE.66.016404} {\bibfield  {journal} {\bibinfo
  {journal} {Phys. Rev. E}\ }\textbf {\bibinfo {volume} {66}},\ \bibinfo
  {pages} {016404} (\bibinfo {year} {2002})}\BibitemShut {NoStop}%
\bibitem [{\citenamefont {Ott}\ \emph {et~al.}(2014)\citenamefont {Ott},
  \citenamefont {Bonitz}, \citenamefont {Stanton},\ and\ \citenamefont
  {Murillo}}]{Ott2014}%
  \BibitemOpen
  \bibfield  {author} {\bibinfo {author} {\bibfnamefont {T.}~\bibnamefont
  {Ott}}, \bibinfo {author} {\bibfnamefont {M.}~\bibnamefont {Bonitz}},
  \bibinfo {author} {\bibfnamefont {L.~G.}\ \bibnamefont {Stanton}}, \ and\
  \bibinfo {author} {\bibfnamefont {M.~S.}\ \bibnamefont {Murillo}},\ }\href
  {\doibase 10.1063/1.4900625} {\bibfield  {journal} {\bibinfo  {journal}
  {Physics of Plasmas}\ }\textbf {\bibinfo {volume} {21}},\ \bibinfo {pages}
  {113704} (\bibinfo {year} {2014})}\BibitemShut {NoStop}%
\bibitem [{Note1()}]{Note1}%
  \BibitemOpen
  \bibinfo {note} {$c_0$ is the conventional dust acoustic speed in a dusty
  plasma}\BibitemShut {NoStop}%
\bibitem [{\citenamefont {Hopkins}\ \emph {et~al.}(2005)\citenamefont
  {Hopkins}, \citenamefont {Archer},\ and\ \citenamefont
  {Evans}}]{Hopkins2005}%
  \BibitemOpen
  \bibfield  {author} {\bibinfo {author} {\bibfnamefont {P.}~\bibnamefont
  {Hopkins}}, \bibinfo {author} {\bibfnamefont {A.~J.}\ \bibnamefont {Archer}},
  \ and\ \bibinfo {author} {\bibfnamefont {R.}~\bibnamefont {Evans}},\ }\href
  {\doibase 10.1103/PhysRevE.71.027401} {\bibfield  {journal} {\bibinfo
  {journal} {Phys. Rev. E}\ }\textbf {\bibinfo {volume} {71}},\ \bibinfo
  {pages} {027401} (\bibinfo {year} {2005})}\BibitemShut {NoStop}%
\bibitem [{\citenamefont {Stringfellow}\ \emph {et~al.}(1990)\citenamefont
  {Stringfellow}, \citenamefont {DeWitt},\ and\ \citenamefont
  {Slattery}}]{Stringfellow1990}%
  \BibitemOpen
  \bibfield  {author} {\bibinfo {author} {\bibfnamefont {G.~S.}\ \bibnamefont
  {Stringfellow}}, \bibinfo {author} {\bibfnamefont {H.~E.}\ \bibnamefont
  {DeWitt}}, \ and\ \bibinfo {author} {\bibfnamefont {W.~L.}\ \bibnamefont
  {Slattery}},\ }\href {\doibase 10.1103/PhysRevA.41.1105} {\bibfield
  {journal} {\bibinfo  {journal} {Phys. Rev. A}\ }\textbf {\bibinfo {volume}
  {41}},\ \bibinfo {pages} {1105} (\bibinfo {year} {1990})}\BibitemShut
  {NoStop}%
\bibitem [{\citenamefont {Hamaguchi}\ \emph {et~al.}(1997)\citenamefont
  {Hamaguchi}, \citenamefont {Farouki},\ and\ \citenamefont
  {Dubin}}]{Hamaguchi1997}%
  \BibitemOpen
  \bibfield  {author} {\bibinfo {author} {\bibfnamefont {S.}~\bibnamefont
  {Hamaguchi}}, \bibinfo {author} {\bibfnamefont {R.~T.}\ \bibnamefont
  {Farouki}}, \ and\ \bibinfo {author} {\bibfnamefont {D.~H.~E.}\ \bibnamefont
  {Dubin}},\ }\href {\doibase 10.1103/PhysRevE.56.4671} {\bibfield  {journal}
  {\bibinfo  {journal} {Phys. Rev. E}\ }\textbf {\bibinfo {volume} {56}},\
  \bibinfo {pages} {4671} (\bibinfo {year} {1997})}\BibitemShut {NoStop}%
\bibitem [{\citenamefont {Khrapak}\ \emph {et~al.}(2004)\citenamefont
  {Khrapak}, \citenamefont {Ivlev},\ and\ \citenamefont
  {Morfill}}]{Khrapak2004}%
  \BibitemOpen
  \bibfield  {author} {\bibinfo {author} {\bibfnamefont {S.~A.}\ \bibnamefont
  {Khrapak}}, \bibinfo {author} {\bibfnamefont {A.~V.}\ \bibnamefont {Ivlev}},
  \ and\ \bibinfo {author} {\bibfnamefont {G.~E.}\ \bibnamefont {Morfill}},\
  }\href {\doibase 10.1103/PhysRevE.70.056405} {\bibfield  {journal} {\bibinfo
  {journal} {Phys. Rev. E}\ }\textbf {\bibinfo {volume} {70}},\ \bibinfo
  {pages} {056405} (\bibinfo {year} {2004})}\BibitemShut {NoStop}%
\bibitem [{\citenamefont {Kampen}(1957)}]{vanKampen1957}%
  \BibitemOpen
  \bibfield  {author} {\bibinfo {author} {\bibfnamefont {N.~V.}\ \bibnamefont
  {Kampen}},\ }\href {\doibase https://doi.org/10.1016/S0031-8914(57)93718-7}
  {\bibfield  {journal} {\bibinfo  {journal} {Physica}\ }\textbf {\bibinfo
  {volume} {23}},\ \bibinfo {pages} {641 } (\bibinfo {year}
  {1957})}\BibitemShut {NoStop}%
\bibitem [{\citenamefont {Pines}\ and\ \citenamefont {Bohm}(1952)}]{Pines1952}%
  \BibitemOpen
  \bibfield  {author} {\bibinfo {author} {\bibfnamefont {D.}~\bibnamefont
  {Pines}}\ and\ \bibinfo {author} {\bibfnamefont {D.}~\bibnamefont {Bohm}},\
  }\href {\doibase 10.1103/PhysRev.85.338} {\bibfield  {journal} {\bibinfo
  {journal} {Phys. Rev.}\ }\textbf {\bibinfo {volume} {85}},\ \bibinfo {pages}
  {338} (\bibinfo {year} {1952})}\BibitemShut {NoStop}%
\bibitem [{\citenamefont {Golden}\ \emph {et~al.}(2018)\citenamefont {Golden},
  \citenamefont {Kalman},\ and\ \citenamefont {Silvestri}}]{Golden2018}%
  \BibitemOpen
  \bibfield  {author} {\bibinfo {author} {\bibfnamefont {K.~I.}\ \bibnamefont
  {Golden}}, \bibinfo {author} {\bibfnamefont {G.~J.}\ \bibnamefont {Kalman}},
  \ and\ \bibinfo {author} {\bibfnamefont {L.~G.}\ \bibnamefont {Silvestri}},\
  }\href {\doibase 10.26577/phst-2017-1-117} {\bibfield  {journal} {\bibinfo
  {journal} {Physical Sciences and Technology}\ }\textbf {\bibinfo {volume}
  {4}},\ \bibinfo {pages} {9} (\bibinfo {year} {2018})}\BibitemShut {NoStop}%
\bibitem [{\citenamefont {Rostoker}\ and\ \citenamefont
  {Rosenbluth}(1960)}]{Rostoker1960}%
  \BibitemOpen
  \bibfield  {author} {\bibinfo {author} {\bibfnamefont {N.}~\bibnamefont
  {Rostoker}}\ and\ \bibinfo {author} {\bibfnamefont {M.~N.}\ \bibnamefont
  {Rosenbluth}},\ }\href {\doibase 10.1063/1.1705998} {\bibfield  {journal}
  {\bibinfo  {journal} {The Physics of Fluids}\ }\textbf {\bibinfo {volume}
  {3}},\ \bibinfo {pages} {1} (\bibinfo {year} {1960})}\BibitemShut {NoStop}%
\bibitem [{\citenamefont {Hartmann}\ \emph {et~al.}(2005)\citenamefont
  {Hartmann}, \citenamefont {Kalman}, \citenamefont {Donk\'o},\ and\
  \citenamefont {Kutasi}}]{Hartmann2005}%
  \BibitemOpen
  \bibfield  {author} {\bibinfo {author} {\bibfnamefont {P.}~\bibnamefont
  {Hartmann}}, \bibinfo {author} {\bibfnamefont {G.~J.}\ \bibnamefont
  {Kalman}}, \bibinfo {author} {\bibfnamefont {Z.}~\bibnamefont {Donk\'o}}, \
  and\ \bibinfo {author} {\bibfnamefont {K.}~\bibnamefont {Kutasi}},\ }\href
  {\doibase 10.1103/PhysRevE.72.026409} {\bibfield  {journal} {\bibinfo
  {journal} {Phys. Rev. E}\ }\textbf {\bibinfo {volume} {72}},\ \bibinfo
  {pages} {026409} (\bibinfo {year} {2005})}\BibitemShut {NoStop}%
\bibitem [{\citenamefont {Rosenfeld}\ and\ \citenamefont
  {Tarazona}(1998)}]{Rosenfeld1998}%
  \BibitemOpen
  \bibfield  {author} {\bibinfo {author} {\bibfnamefont {Y.}~\bibnamefont
  {Rosenfeld}}\ and\ \bibinfo {author} {\bibfnamefont {P.}~\bibnamefont
  {Tarazona}},\ }\href {\doibase 10.1080/00268979809483145} {\bibfield
  {journal} {\bibinfo  {journal} {Molecular Physics}\ }\textbf {\bibinfo
  {volume} {95}},\ \bibinfo {pages} {141} (\bibinfo {year} {1998})},\ \Eprint
  {http://arxiv.org/abs/https://doi.org/10.1080/00268979809483145}
  {https://doi.org/10.1080/00268979809483145} \BibitemShut {NoStop}%
\bibitem [{\citenamefont {Rosenfeld}(2000)}]{Rosenfeld2000}%
  \BibitemOpen
  \bibfield  {author} {\bibinfo {author} {\bibfnamefont {Y.}~\bibnamefont
  {Rosenfeld}},\ }\href {\doibase 10.1103/PhysRevE.62.7524} {\bibfield
  {journal} {\bibinfo  {journal} {Phys. Rev. E}\ }\textbf {\bibinfo {volume}
  {62}},\ \bibinfo {pages} {7524} (\bibinfo {year} {2000})}\BibitemShut
  {NoStop}%
\bibitem [{\citenamefont {Khrapak}\ and\ \citenamefont
  {Thomas}(2015{\natexlab{b}})}]{Khrapak2015a}%
  \BibitemOpen
  \bibfield  {author} {\bibinfo {author} {\bibfnamefont {S.~A.}\ \bibnamefont
  {Khrapak}}\ and\ \bibinfo {author} {\bibfnamefont {H.~M.}\ \bibnamefont
  {Thomas}},\ }\href {\doibase 10.1103/PhysRevE.91.023108} {\bibfield
  {journal} {\bibinfo  {journal} {Phys. Rev. E}\ }\textbf {\bibinfo {volume}
  {91}},\ \bibinfo {pages} {023108} (\bibinfo {year}
  {2015}{\natexlab{b}})}\BibitemShut {NoStop}%
\bibitem [{\citenamefont {Hansen}\ and\ \citenamefont
  {McDonald}(2013)}]{HansenBook}%
  \BibitemOpen
  \bibfield  {author} {\bibinfo {author} {\bibfnamefont {J.-P.}\ \bibnamefont
  {Hansen}}\ and\ \bibinfo {author} {\bibfnamefont {I.~R.}\ \bibnamefont
  {McDonald}},\ }in\ \href {\doibase
  https://doi.org/10.1016/B978-0-12-387032-2.00008-8} {\emph {\bibinfo
  {booktitle} {Theory of Simple Liquids (Fourth Edition)}}}\ (\bibinfo
  {publisher} {Academic Press},\ \bibinfo {address} {Oxford},\ \bibinfo {year}
  {2013})\ \bibinfo {edition} {fourth edition}\ ed.,\ pp.\ \bibinfo {pages}
  {311 -- 361}\BibitemShut {NoStop}%
\bibitem [{\citenamefont {Golden}\ and\ \citenamefont
  {Kalman}(2003)}]{Golden2003}%
  \BibitemOpen
  \bibfield  {author} {\bibinfo {author} {\bibfnamefont {K.~I.}\ \bibnamefont
  {Golden}}\ and\ \bibinfo {author} {\bibfnamefont {G.~J.}\ \bibnamefont
  {Kalman}},\ }\href {\doibase 10.1088/0305-4470/36/22/306} {\bibfield
  {journal} {\bibinfo  {journal} {Journal of Physics A: Mathematical and
  General}\ }\textbf {\bibinfo {volume} {36}},\ \bibinfo {pages} {5865}
  (\bibinfo {year} {2003})}\BibitemShut {NoStop}%
\bibitem [{\citenamefont {Feynman}(1954)}]{Feynman1954}%
  \BibitemOpen
  \bibfield  {author} {\bibinfo {author} {\bibfnamefont {R.~P.}\ \bibnamefont
  {Feynman}},\ }\href {\doibase 10.1103/PhysRev.94.262} {\bibfield  {journal}
  {\bibinfo  {journal} {Phys. Rev.}\ }\textbf {\bibinfo {volume} {94}},\
  \bibinfo {pages} {262} (\bibinfo {year} {1954})}\BibitemShut {NoStop}%
\bibitem [{\citenamefont {Feynman}\ and\ \citenamefont
  {Cohen}(1956)}]{Feynman1956}%
  \BibitemOpen
  \bibfield  {author} {\bibinfo {author} {\bibfnamefont {R.~P.}\ \bibnamefont
  {Feynman}}\ and\ \bibinfo {author} {\bibfnamefont {M.}~\bibnamefont
  {Cohen}},\ }\href {\doibase 10.1103/PhysRev.102.1189} {\bibfield  {journal}
  {\bibinfo  {journal} {Phys. Rev.}\ }\textbf {\bibinfo {volume} {102}},\
  \bibinfo {pages} {1189} (\bibinfo {year} {1956})}\BibitemShut {NoStop}%
\bibitem [{\citenamefont {Cohen}\ and\ \citenamefont
  {Feynman}(1957)}]{Cohen1957}%
  \BibitemOpen
  \bibfield  {author} {\bibinfo {author} {\bibfnamefont {M.}~\bibnamefont
  {Cohen}}\ and\ \bibinfo {author} {\bibfnamefont {R.~P.}\ \bibnamefont
  {Feynman}},\ }\href {\doibase 10.1103/PhysRev.107.13} {\bibfield  {journal}
  {\bibinfo  {journal} {Phys. Rev.}\ }\textbf {\bibinfo {volume} {107}},\
  \bibinfo {pages} {13} (\bibinfo {year} {1957})}\BibitemShut {NoStop}%
\bibitem [{\citenamefont {{Goree}}\ and\ \citenamefont
  {{Ruhunusiri}}(2014)}]{Goree2014}%
  \BibitemOpen
  \bibfield  {author} {\bibinfo {author} {\bibfnamefont {J.}~\bibnamefont
  {{Goree}}}\ and\ \bibinfo {author} {\bibfnamefont {W.~D.~S.}\ \bibnamefont
  {{Ruhunusiri}}},\ }in\ \href
  {http://meetings.aps.org/link/BAPS.2014.DPP.BO3.2} {\emph {\bibinfo
  {booktitle} {APS Meeting Abstracts}}}\ (\bibinfo {year} {2014})\ p.\ \bibinfo
  {pages} {BO3.002}\BibitemShut {NoStop}%
\bibitem [{\citenamefont {Kalman}\ and\ \citenamefont
  {Golden}(1990)}]{QLCA1990}%
  \BibitemOpen
  \bibfield  {author} {\bibinfo {author} {\bibfnamefont {G.~J.}\ \bibnamefont
  {Kalman}}\ and\ \bibinfo {author} {\bibfnamefont {K.~I.}\ \bibnamefont
  {Golden}},\ }\href {\doibase 10.1103/PhysRevA.41.5516} {\bibfield  {journal}
  {\bibinfo  {journal} {Phys. Rev. A}\ }\textbf {\bibinfo {volume} {41}},\
  \bibinfo {pages} {5516} (\bibinfo {year} {1990})}\BibitemShut {NoStop}%
\bibitem [{\citenamefont {Golden}\ and\ \citenamefont
  {Kalman}(2000)}]{QLCA2000}%
  \BibitemOpen
  \bibfield  {author} {\bibinfo {author} {\bibfnamefont {K.~I.}\ \bibnamefont
  {Golden}}\ and\ \bibinfo {author} {\bibfnamefont {G.~J.}\ \bibnamefont
  {Kalman}},\ }\href {\doibase 10.1063/1.873814} {\bibfield  {journal}
  {\bibinfo  {journal} {Phys. Plasmas}\ }\textbf {\bibinfo {volume} {7}},\
  \bibinfo {pages} {14} (\bibinfo {year} {2000})}\BibitemShut {NoStop}%
\bibitem [{\citenamefont {Golden}\ and\ \citenamefont
  {Kalman}(2001)}]{QLCA2000Err}%
  \BibitemOpen
  \bibfield  {author} {\bibinfo {author} {\bibfnamefont {K.~I.}\ \bibnamefont
  {Golden}}\ and\ \bibinfo {author} {\bibfnamefont {G.~J.}\ \bibnamefont
  {Kalman}},\ }\href {\doibase 10.1063/1.1408622} {\bibfield  {journal}
  {\bibinfo  {journal} {Physics of Plasmas}\ }\textbf {\bibinfo {volume} {8}},\
  \bibinfo {pages} {5064} (\bibinfo {year} {2001})}\BibitemShut {NoStop}%
\bibitem [{Note2()}]{Note2}%
  \BibitemOpen
  \bibinfo {note} {The factor $4\pi q^2$ in the denominator of the second term
  is due to our definition of $\phi (r) \propto q^2$}\BibitemShut {NoStop}%
\bibitem [{{\relax DLMF}()}]{NIST:DLMF}%
  \BibitemOpen
  {\relax DLMF},\ \href {http://dlmf.nist.gov/} {\enquote {\bibinfo {title}
  {{\it NIST Digital Library of Mathematical Functions}},}\ }\bibinfo
  {howpublished} {http://dlmf.nist.gov/, Release 1.0.22 of 2019-03-15},\
  \bibinfo {note} {f.~W.~J. Olver, A.~B. {Olde Daalhuis}, D.~W. Lozier, B.~I.
  Schneider, R.~F. Boisvert, C.~W. Clark, B.~R. Miller and B.~V. Saunders,
  eds.}\BibitemShut {Stop}%
\bibitem [{\citenamefont {Fried}\ and\ \citenamefont
  {Conte}(1961)}]{Fried1961}%
  \BibitemOpen
  \bibfield  {author} {\bibinfo {author} {\bibfnamefont {B.~D.}\ \bibnamefont
  {Fried}}\ and\ \bibinfo {author} {\bibfnamefont {S.~D.}\ \bibnamefont
  {Conte}},\ }\href@noop {} {\emph {\bibinfo {title} {The Plasma Dispersion
  Function: {T}he {H}ilbert Transform of the {G}aussian.}}}\ (\bibinfo
  {publisher} {Academic Press},\ \bibinfo {address} {London-New York},\
  \bibinfo {year} {1961})\ pp.\ \bibinfo {pages} {v+419},\ \bibinfo {note}
  {erratum: Math. Comp. v. 26 (1972), no. 119, p. 814.}\BibitemShut {Stop}%
\bibitem [{\citenamefont {Faddeyeva}\ and\ \citenamefont
  {Terent'ev}(1961)}]{Faddeyeva1961}%
  \BibitemOpen
  \bibfield  {author} {\bibinfo {author} {\bibfnamefont {V.~N.}\ \bibnamefont
  {Faddeyeva}}\ and\ \bibinfo {author} {\bibfnamefont {N.~M.}\ \bibnamefont
  {Terent'ev}},\ }\href@noop {} {\emph {\bibinfo {title} {Tables of Values of
  the Function {$w(z)=e^{-z^{2}}(1+2i\pi^{-1/2}\int_{0}^{z}e^{t^{2}}dt)$} for
  Complex Argument}}},\ Edited by V. A. Fok; translated from the Russian by D.
  G. Fry. Mathematical Tables Series, Vol. 11\ (\bibinfo  {publisher} {Pergamon
  Press},\ \bibinfo {address} {Oxford},\ \bibinfo {year} {1961})\ pp.\ \bibinfo
  {pages} {v+280}\BibitemShut {NoStop}%
\bibitem [{\citenamefont {Chapman}\ \emph {et~al.}(1990)\citenamefont
  {Chapman}, \citenamefont {Cowling}, \citenamefont {Burnett},\ and\
  \citenamefont {Cercignani}}]{Chapman1990}%
  \BibitemOpen
  \bibfield  {author} {\bibinfo {author} {\bibfnamefont {S.}~\bibnamefont
  {Chapman}}, \bibinfo {author} {\bibfnamefont {T.}~\bibnamefont {Cowling}},
  \bibinfo {author} {\bibfnamefont {D.}~\bibnamefont {Burnett}}, \ and\
  \bibinfo {author} {\bibfnamefont {C.}~\bibnamefont {Cercignani}},\
  }\href@noop {} {\emph {\bibinfo {title} {{The Mathematical Theory of
  Non-uniform Gases: An Account of the Kinetic Theory of Viscosity, Thermal
  Conduction and Diffusion in Gases}}}},\ Cambridge Mathematical Library\
  (\bibinfo  {publisher} {Cambridge University Press},\ \bibinfo {year}
  {1990})\BibitemShut {NoStop}%
\bibitem [{\citenamefont {Oberman}\ \emph {et~al.}(1962)\citenamefont
  {Oberman}, \citenamefont {Ron},\ and\ \citenamefont {Dawson}}]{Oberman1962}%
  \BibitemOpen
  \bibfield  {author} {\bibinfo {author} {\bibfnamefont {C.}~\bibnamefont
  {Oberman}}, \bibinfo {author} {\bibfnamefont {A.}~\bibnamefont {Ron}}, \ and\
  \bibinfo {author} {\bibfnamefont {J.}~\bibnamefont {Dawson}},\ }\href
  {\doibase 10.1063/1.1706560} {\bibfield  {journal} {\bibinfo  {journal}
  {Physics of Fluids}\ }\textbf {\bibinfo {volume} {5}},\ \bibinfo {pages}
  {1514} (\bibinfo {year} {1962})}\BibitemShut {NoStop}%
\bibitem [{\citenamefont {DuBois}\ \emph {et~al.}(1962)\citenamefont {DuBois},
  \citenamefont {Gilinsky},\ and\ \citenamefont {Kivelson}}]{DuBois1962}%
  \BibitemOpen
  \bibfield  {author} {\bibinfo {author} {\bibfnamefont {D.~F.}\ \bibnamefont
  {DuBois}}, \bibinfo {author} {\bibfnamefont {V.}~\bibnamefont {Gilinsky}}, \
  and\ \bibinfo {author} {\bibfnamefont {M.~G.}\ \bibnamefont {Kivelson}},\
  }\href {\doibase 10.1103/PhysRevLett.8.419} {\bibfield  {journal} {\bibinfo
  {journal} {Physical Review Letters}\ }\textbf {\bibinfo {volume} {8}},\
  \bibinfo {pages} {419} (\bibinfo {year} {1962})}\BibitemShut {NoStop}%
\bibitem [{\citenamefont {Coste}(1965)}]{Coste1965b}%
  \BibitemOpen
  \bibfield  {author} {\bibinfo {author} {\bibfnamefont {J.}~\bibnamefont
  {Coste}},\ }\href {\doibase 10.1088/0029-5515/5/4/006} {\bibfield  {journal}
  {\bibinfo  {journal} {Nuclear Fusion}\ }\textbf {\bibinfo {volume} {5}},\
  \bibinfo {pages} {293} (\bibinfo {year} {1965})}\BibitemShut {NoStop}%
\end{thebibliography}
\end{document}